\documentclass[%
reprint,
superscriptaddress,
preprintnumbers,
 amsmath,amssymb,
 aps,
pra,
]{revtex4-2}
\usepackage{multirow}
\usepackage{subfigure}
\usepackage{graphicx}
\usepackage{dcolumn}
\usepackage{bm}
\usepackage{color}
\usepackage{dsfont}
\usepackage[normalem]{ulem}
\usepackage{hyperref}
\hypersetup{colorlinks=true,allcolors=blue}
\usepackage{wasysym}
\newcommand\tinyvarhexagon{\vcenter{\hbox{\scalebox{0.7}{$\varhexagon$}}}}

\begin{document}
\title{Transport in honeycomb lattice with random $\pi$-fluxes: Implications for low-temperature thermal transport in the Kitaev spin liquids}
\author{Zekun Zhuang}
\affiliation{Center for Materials Theory, Rutgers University, Piscataway, New Jersey 08854, USA}


\begin{abstract}
Motivated by the thermal transport problem in the Kitaev spin liquids, we consider a nearest-neighbor tight-binding model on the honeycomb lattice in the presence of random uncorrelated $\pi$-fluxes. We employ different numerical methods to study its transport properties near half-filling. The zero-temperature DC conductivity away from the Dirac point is found to be quadratic in Fermi momentum and inversely proportional to the flux density. Localization due to the random $\pi$-fluxes is observed and the localization length is extracted. Our results imply that, for realistic system size, the thermal conductivity of a pure Kitaev spin liquid diverges as $\kappa_\text{K}\sim T^3 e^{\Delta_v/k_BT}$ when $k_B T\ll \Delta_v$, and suggest the possible occurrence of strong Majorana localization $\kappa_\text{K}/T\ll k_B^2/2\pi\hbar$ when $k_B T\sim \Delta_v$, where $\Delta_v$ is the vison gap. 
\end{abstract}

\maketitle



\section{Introduction}
Quantum spin liquids (QSLs) are an exotic state of matter with no local symmetry breaking, deconfined fractionalized quasiparticles, and emergent gauge excitations \cite{Anderson1973,Savary2016,Zhou2017}. The Kitaev model, an exactly solvable honeycomb model, provides a typical framework for describing such a phase, where spins fractionalize to Majorana fermions and $Z_2$ gauge excitations (visons) \cite{Kitaev2006}. Thanks to the proposal by Jackeli and Khaliullin \cite{Jackeli2009}, which suggests that the Kitaev model can be realized in certain strongly spin-orbit coupled systems, a growing number of Kitaev candidate materials have been discovered in the past decade \cite{Winter2017,Trebst2022}.

Despite the theoretical proposal of QSLs for several decades, their experimental identification remains challenging. Thermal transport experiments are a promising technique for characterizing QSLs, as they allow the detection of charge-neutral, mobile quasiparticles. For example, in the Kitaev candidate material $\alpha-\text{RuCl}_3$, a half-quantized thermal Hall conductance has been observed, indicating the existence of a chiral Majorana mode at the edge of the non-Abelian Kitaev spin liquid \cite{Kasahara2018,Kasahara2018_2,Yokoi2021}. To fully understand the experimental results, it is crucial to predict the thermal transport signatures of QSLs, particularly those proximate to the exact Kitaev model, in various different regimes \cite{Nasu2017,Metavitsiadis2017,Pidatella2019,Kao2021,Nasu2020,Joy2022,Zheng2021,Chen2023,Koyama2021,Zhang2021,Cookmeyer2018}.
 
The longitudinal thermal conductivity $\kappa_{\text{xx}}(\omega,T)$ of the Kitaev model has been numerically investigated using Kubo's formalism, and the DC thermal conductivity $\kappa_\text{K}(T)$ can be obtained by extrapolation of $\kappa_{\text{xx}}(\omega,T)$ to the $\omega\rightarrow 0$ limit \cite{Nasu2017,Metavitsiadis2017,Pidatella2019,Kao2021,Nasu2020}. When the temperature $T$ is comparable to or smaller than the vison gap $\Delta_v$, such extrapolation appears to give results with significant error bars and calculations with larger system sizes are needed \cite{Nasu2017,Nasu2020}, leaving the low-temperature behavior of $\kappa_\text{K}$ inconclusive. Neglecting the gauge excitations in the Kitaev model, it is possible to regard an undoped graphene as two stacks of the Kitaev model, and thus expect that the thermal transport of Kitaev spin liquid and the electric transport of graphene may have similar low-temperature behavior. It was proposed that in graphene there exists a universal minimum electrical conductivity $\sigma_\text{min}=e^2/\pi h$ per valley per spin, regardless of the concentration of disorders \cite{Fradkin1986,Lee1993,Peres2006}. It is thus natural to ask whether a similar `minimum thermal conductivity' (MTC) also exists in a pure Kitaev model. A previous quantum Monte Carlo study seems to support the existence of MTC in the low-temperature regime, i.e. $\lim_{T\rightarrow 0} \kappa_\text{K}/T=k_B^2/12\hbar$, but further justification is required due to the significant uncertainties and finite-size effects in the numerical simulation \cite{Nasu2017}. Moreover, at low temperatures, the phase coherence length may exceed the dimensions of a realistic sample, within which the quantum interference effects are essential. Consequently, the thermally excited random $Z_2$ flux background may strongly localize Majorana particles \cite{Zhu2021,Kao2021,Kim2022}, provided that the localization length is smaller than the system size. Such localization effects are uncaptured by previous calculations performed for relatively small systems, and overlooking these effects may crucially affect the interpretation of experimental results.

\begin{figure}[b]  
\subfigure[]{
    \label{fig:Kitaev} 
    \includegraphics[height=1.4 in]{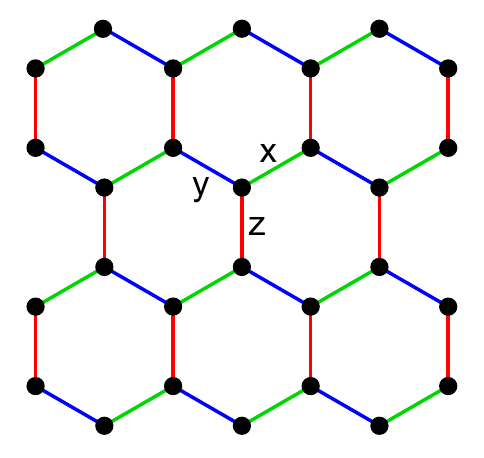}}  
\subfigure[]{
    \label{fig:fermionmodel} 
    \includegraphics[height=1.4 in]{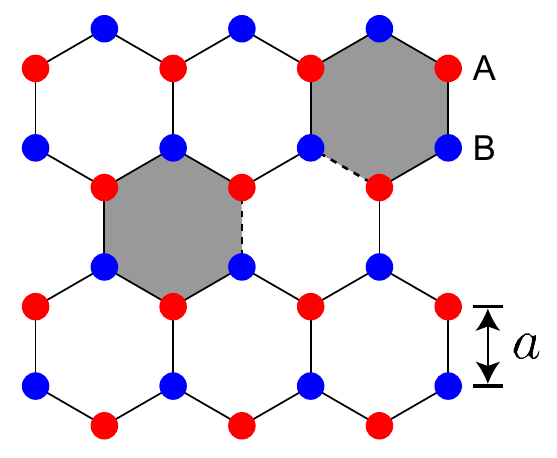}}

    \caption{(a) Kitaev on the honeycomb lattice, where different colors denote different types of bonds. (b) The nearest-neighbor tight-binding model on the honeycomb lattice threaded by random $\pi$-fluxes (grey areas), where the solid (dashed) lines denote the bonds on which the $Z_2$ gauge field takes value $1(-1)$.  }
    \label{fig:Lattice}
\end{figure}

To understand the low-temperature thermal conductivity of the Kitaev model in the absence of disorder and magnetic fields, in this work we numerically study the transport of complex fermions in a random $\pi$-flux honeycomb model (RPFHM) near half-filling. This approach is motivated by the observation that the Kitaev model (see Fig. \ref{fig:Kitaev}) can be seen as half of the RPFHM (see Fig. \ref{fig:fermionmodel}), and the thermal conductivity of the former is linked to the electrical conductivity of the latter via the Wiedemann-Franz (WF) law. Furthermore, the RPFHM is of academic interest on its own, as this random $Z_2$-flux problem is less studied compared to random $U(1)$-flux models \cite{Ohtsuki1993,Tadjine2018,Hart2020,Hart2021,Li2022}.

We focus on the zero-temperature longitudinal DC conductivity of the RPFHM and its localization properties as a function of flux density and chemical potential. We extract the dependence of DC conductivity on flux density and Fermi wavevector in the semi-classical diffusive regime. Our analysis reveals Anderson localization due to random $\pi$-fluxes, and we obtain 2-D localization lengths. Our results indicate that the low-temperature thermal conductivity of a clean Kitaev model neither exhibits MTC nor vanishes, but instead diverges as $\kappa_\text{K}\sim T^3 e^{\Delta_v/k_BT}$, which is unexpected based on previous studies. Additionally, when $k_BT\sim \Delta_v$, the thermal conductivity could be significantly suppressed due to strong Majorana localization.

\section{Random $\pi$-flux honeycomb model}
\subsection{Model}
We consider a spinless tight-binding model on the honeycomb lattice (Fig. \ref{fig:fermionmodel}) with Hamiltonian
 \begin{equation}
    H_\text{G}=-t\sum_{\langle i,j\rangle} u_{ij} f_i^\dagger f_j, \label{RPFHM}
\end{equation}
where $\langle i,j\rangle$ denotes the nearest-neighbor coupling between site $i$ and $j$, $t$ is a real number and $u_{ij}=u_{ji}=\pm 1$ is the $Z_2$ gauge field coupled to charge $e$ fermions. The creation and annihilation operators obey the anticommutation relation $\{f_i,f_j^\dagger\}=\delta_{ij}$. To avoid the gauge redundancy, one could define the gauge-invariant $Z_2$ flux operator $W_p=\prod_{\langle jk\rangle\in p} u_{jk}=\pm 1$ on each plaquette $p$. On each plaquette, the value of $Z_2$ flux takes $1$ with probability $1-n_v$ and $-1$ ($\pi$-flux) with probability $n_v$, and $W_p$ on different hexagons are uncorrelated. We emphasize that this differs from having independent random $Z_2$ gauge fields on each bond. According to the Altland-Zirnbauer classification \cite{Altland1997,Evers2008,Ludwig2016}, the system is in the orthogonal symmetry class AI when $\epsilon=E/t\neq 0$, while it belongs to the chiral orthogonal symmetry class BDI at the Dirac point $\epsilon=0$.  We aim to calculate the quench average of gauge-invariant observables, which are demonstrated in the following.

\begin{figure}[ht]
    \centering
    \includegraphics[width=0.48\textwidth]{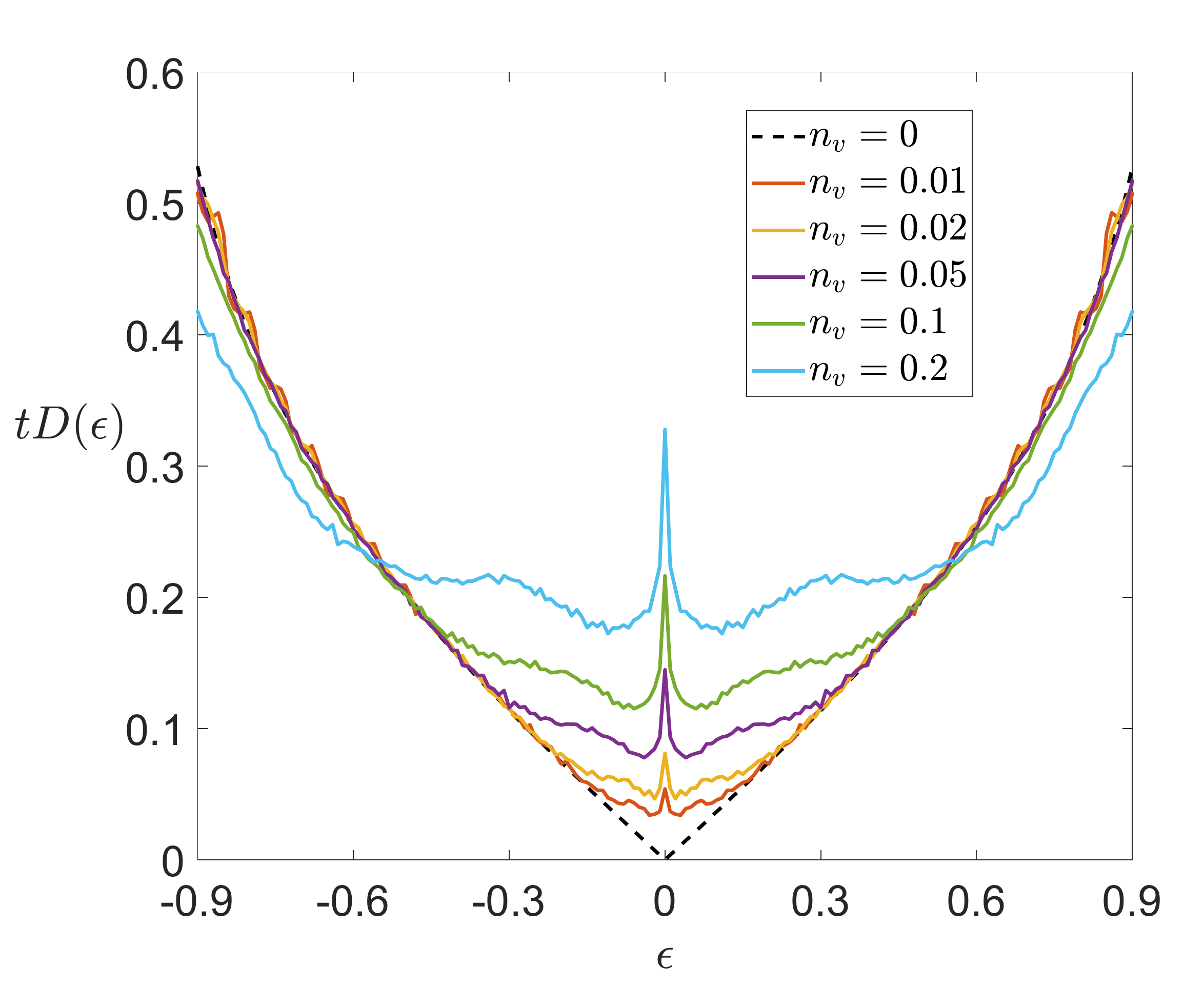}
    \caption{
    The averaged DOS near the Dirac point for different flux densities. The DOS for $n_v\neq 0$ is obtained via exact diagonalization of 10000-site systems and averaging over 190 random flux configurations.}
    \label{fig:DOS}
\end{figure}
\subsection{Density of states}
In the fluxless sector, Eq. (\ref{RPFHM}) corresponds to the nearest-neighbor tight-binding model of graphene, which can be easily diagonalized and the analytic form of its density of states (DOS) is well known \cite{Castro2009}. When $\pi$ fluxes are present, as seen in Fig. \ref{fig:DOS}, for $n_v\lesssim 0.1$ the DOS close to the Dirac point is greatly enhanced, while the DOS away from the Dirac point but below the Van Hove point remains almost unaffected. Remarkably, there appears a sharp DOS peak at the Dirac point, which is a characteristic of the chiral orthogonal symmetry class BDI \cite{Gade1991,Gade1993,Hafner2014,Ferreira2015,Sanyal2016,Ostrovsky2014}. The non-vanishing DOS around the Dirac point suggests that, as long as the low-energy states are not fully localized, the conductance will be finite around the Dirac point and might even be enhanced due to the increasing DOS \cite{Titov2007}. This contrasts with a pristine graphene whose conductance is either zero or of a few conductance quantum $e^2/h$ \cite{Brey2006,Peres2006_2}.
\begin{figure}[b]
    \centering
    \includegraphics[width=0.49\textwidth]{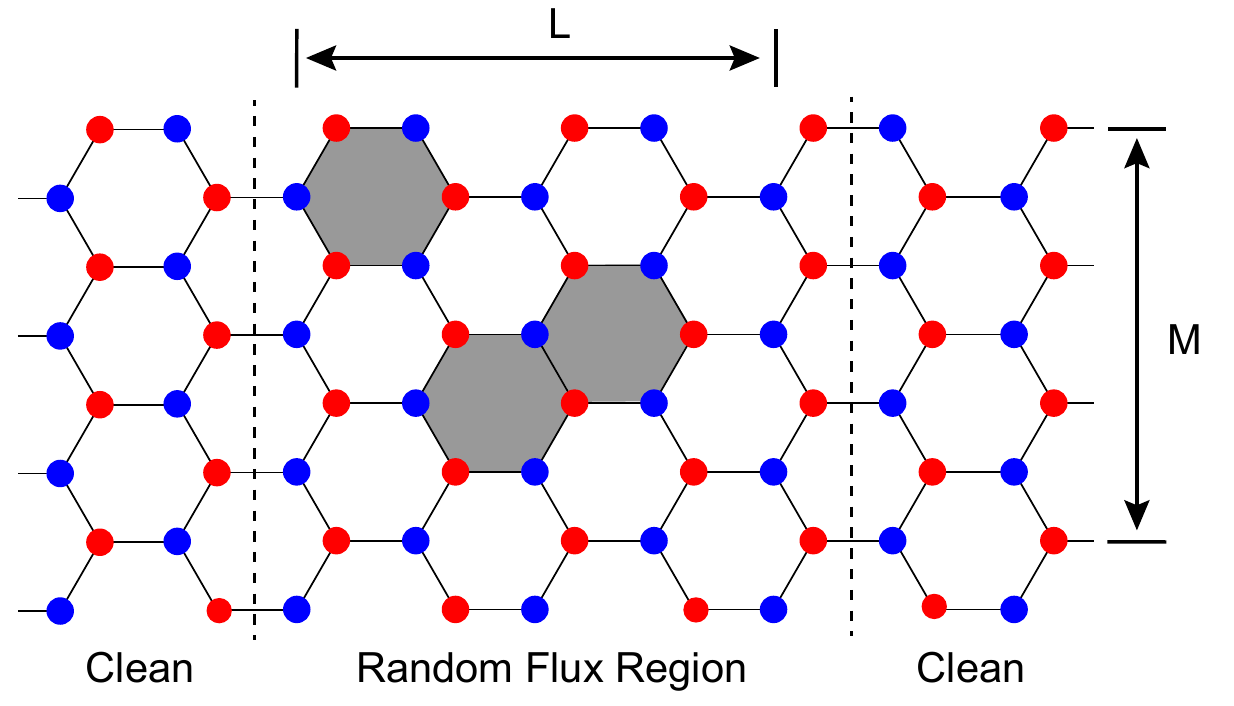}
    \caption{Schematics of the two-terminal setup used to compute the DC conductivity of RPFHM. The middle region, which is penetrated by random $\pi$-fluxes (grey areas), is sandwiched between two semi-infinite `clean' leads.}
    \label{fig:setup}
\end{figure}

\subsection{DC conductivity}
\begin{figure}[!ht]
  \subfigure[]{
    \label{fig:Conductivity1} 
    \includegraphics[width=0.4 \textwidth]{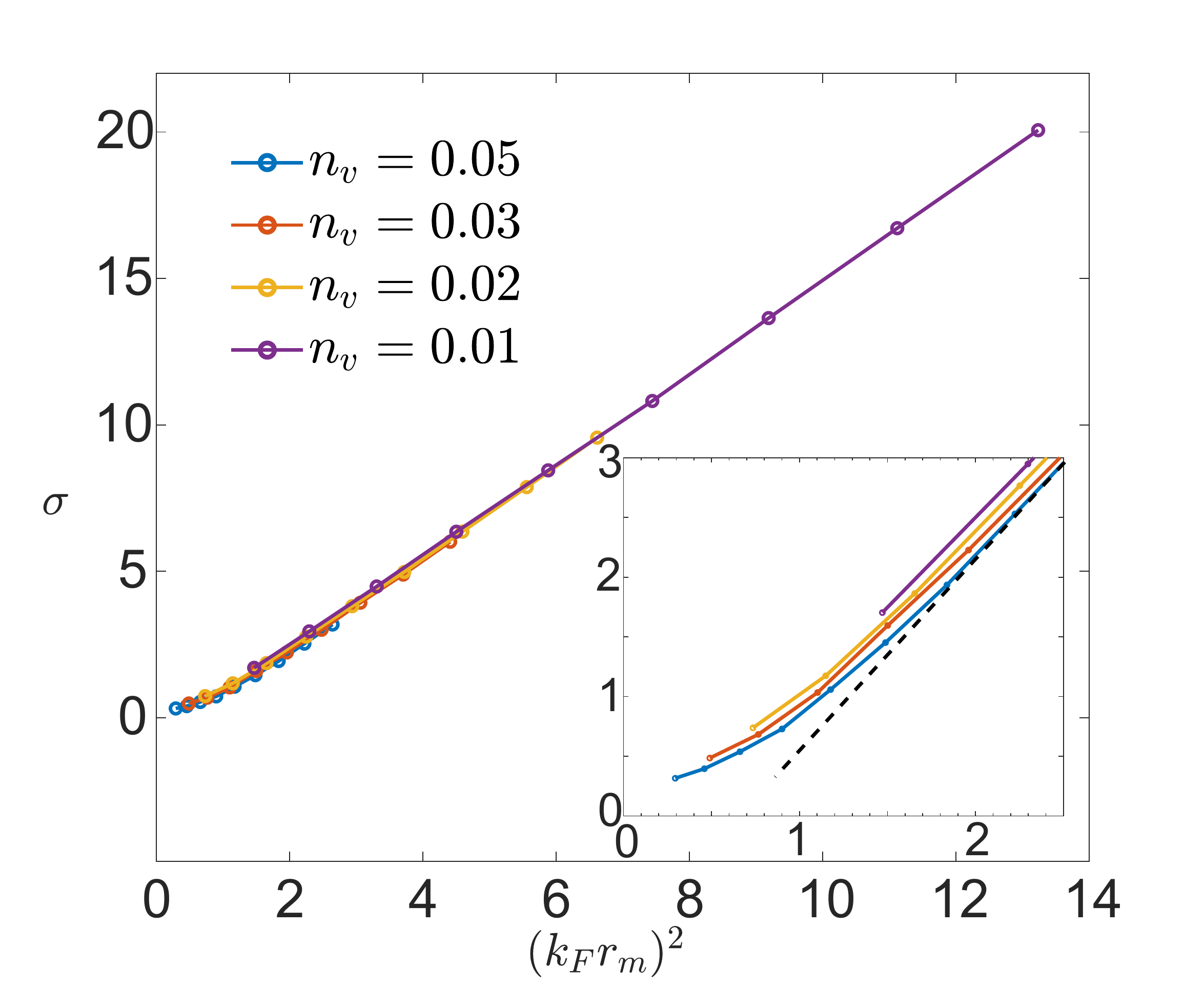}}
    \subfigure[]{
    \label{fig:Conductivity2} 
    \includegraphics[width=0.4 \textwidth]{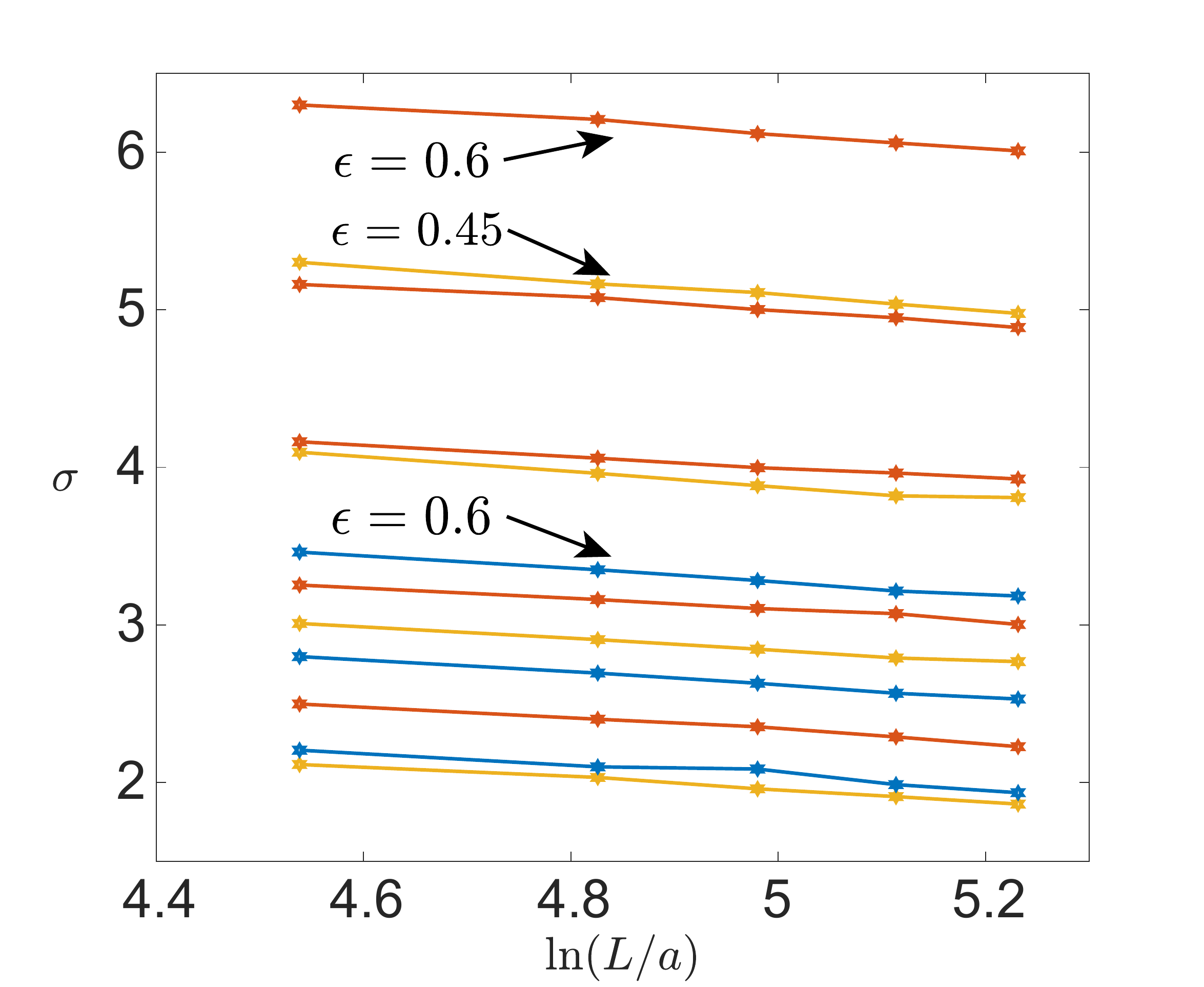}}
    \caption{(a) DC conductivity $\sigma_\text{G}$ (in unit of $e^2/h$) versus $(k_F r_m)^2$ for different $n_v$ when $L\approx M=108\sqrt{3}a$; the inset is the zoom-in plot. (b) DC conductivity $\sigma_\text{G}$ versus $\ln(L/a)$ for different energies and flux densities $n_v=0.02 (\text{yellow}),0.03 (\text{red}),0.05 (\text{blue})$ and $\epsilon$; from top to the bottom, $\epsilon$ decreases by 0.05 for same $n_v$. The error bars are smaller than the marker size and hence not shown.}
    \label{fig:Conductivity}
\end{figure}To compute the DC conductivity, we employ the recursive Green's function method to calculate the transmission function $\mathbf{T}(E,M,L)$ between the left and right leads depicted in Fig. \ref{fig:setup}, as a function of energy $E$, width $M$ and length $L$. The conductance is given by the Landauer formula $G=\mathbf{T}e^2/h$ \cite{Landauer1992}, and the conductivity can be extracted by subtracting the contact resistance from the total resistance
\begin{equation}
    \sigma_\text{G}=\frac{GL}{M}=\frac{L}{M}\frac{1}{\mathbf{T}(E,M,L)^{-1}-\mathbf{T}(E,M,0)^{-1}}\frac{e^2}{h}. \label{DCconductivity}
\end{equation}
The transmission coefficient can be calculated by \cite{Fisher1981,Meir1992}
\begin{equation}
    \mathbf{T}=\text{Tr}(\Gamma_L G^R \Gamma_R G^A), \label{transmissionC}
\end{equation}
where $G^{R(A)}=(E-H_\text{G}-\Sigma_L^{R(A)}-\Sigma_R^{R(A)})^{-1}$ is the retarded (advanced) Green's function in the presence of semi-infinite leads, $\Sigma_\alpha^{R(A)}$ is the self-energy due to the lead $\alpha$, and $\Gamma_\alpha=i(\Sigma_\alpha^R-\Sigma_\alpha^A)$. The self-energies of the leads can be obtained by computing the transfer matrix iteratively \cite{Sancho1984,Nardelli1999} and the bulk Green's function can be calculated efficiently with recursive method \cite{Mackinnon1985}.
Although we only consider the case that the edge is armchair-type, we expect the result for the zigzag edge to be similar in the thermodynamic limit, as the pure system is isotropic at low energies. We compute the conductivity of 2000 different samples and then take the mean value. In general, the conductivity depends on the aspect ratio $M/L$ and we fix it to be close to unity.

Figure \ref{fig:Conductivity1} shows the dependence of DC conductivity on the dimensionless parameter $k_F r_m$ for fixed system size $L=108\sqrt{3}a$, where $r_m=(3\sqrt{3}a^2/2\pi n_v)^{1/2}$ is the average distance between $\pi$-fluxes. Due to the existence of particle-hole symmetry, we only show results for positive $\epsilon$ hereafter. We also only demonstrate the results away from the Dirac point where the DOS is not significantly changed by $\pi$-fluxes, such that Eq. (\ref{DCconductivity}) is applicable and the conductivity is not underestimated due to the vanishing DOS of leads \cite{Croy2006}. The failure of the Landauer approach at the Dirac point can also be understood by noting that it is based on the calculation of transition probabilities between different unperturbed states in clean leads, while the unperturbed eigenstates may not be a good description in the vicinity of the Dirac point where the disorder effects are significant. When the Fermi wavelength $\lambda_F=2\pi/k_F$ of Dirac fermion is much shorter than the mean flux distance $r_m$, one observes that the DC conductivity is approximately linear in $(k_F r_m)^2$, namely
\begin{equation}
  \sigma_\text{G}/(e^2/h)\approx d (k_F r_m)^2+b,  \label{DCconductivity2}
\end{equation} 
where $d\approx 1.6$, and $b$ weakly depends on $n_v$ and $L$ (see also Fig. \ref{fig:Conductivity2}). In this work, we assume the above relation holds as long as the momentum $k_F$ is small so that the dispersion can be regarded as linear. Compared to the Drude conductivity $\sigma_\text{sc}=k_F l_e e^2/h$, this suggests a semiclassical transport regime where, for sufficiently large $\sigma_\text{G}$, the mean free path $l_e$ is approximately inversely proportional to $n_v$ and proportional to the Fermi wavevector $k_F$. Using the Drude formula, we estimate that in Fig. \ref{fig:Conductivity1}, the largest mean free path $l_{e,\text{max}}\approx 80a$ is smaller than the system size $L$, consistent with the diffusive transport regime for which the Drude formula is applicable. When $\sigma_\text{G}\lesssim e^2/h$, which also coincides with $k_F r_m\lesssim 1$, the system enters the quantum regime where the conductivity starts to saturate, evidenced by the upturn at small $k_F r_m$ shown in the inset of Fig. \ref{fig:Conductivity1}. 

\begin{figure*}[!ht]
    \centering
      \subfigure[]{
    \label{fig:localization1} 
    \includegraphics[width=0.32\textwidth]{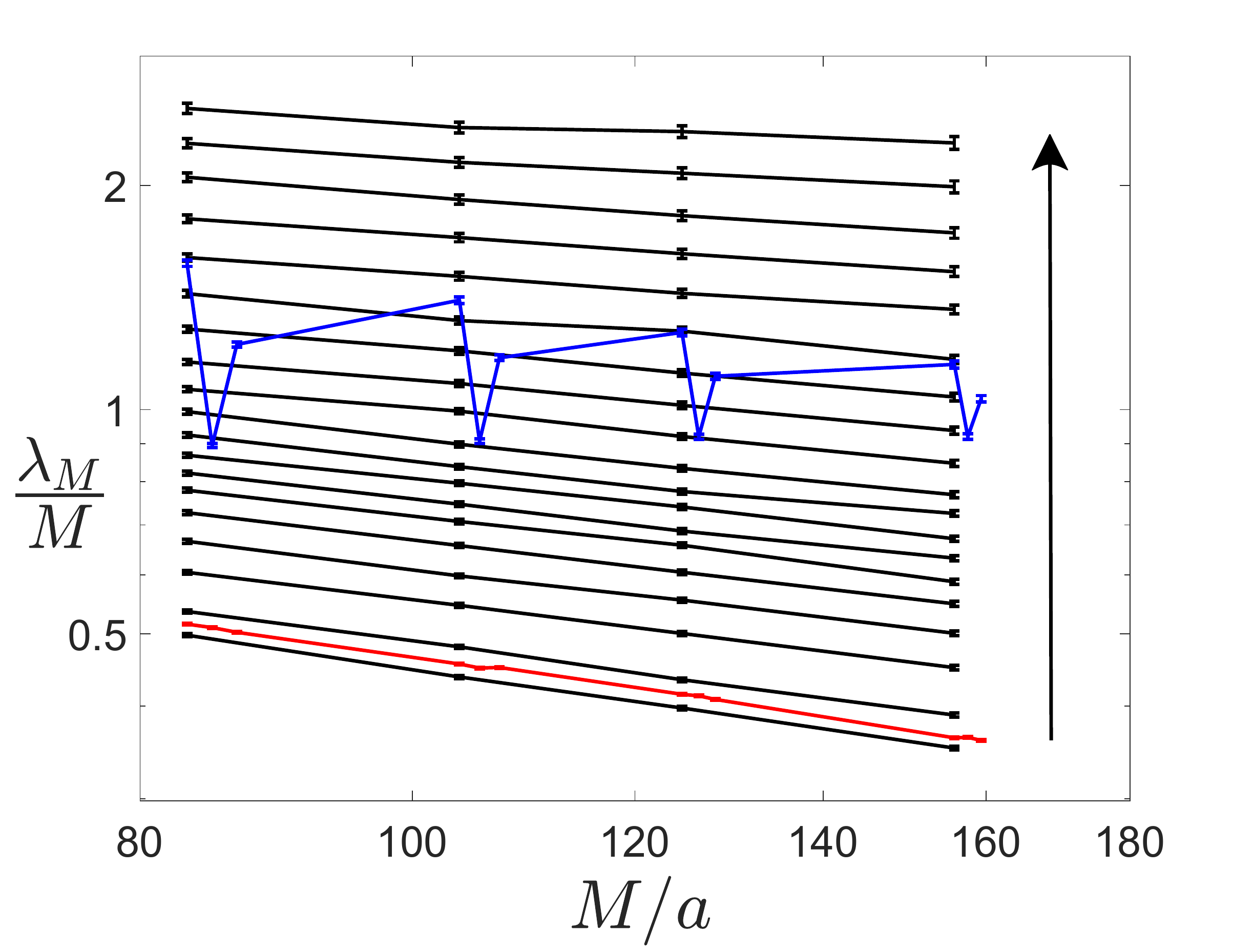}}
    \subfigure[]{
    \label{fig:localization2} 
    \includegraphics[width=0.32\textwidth]{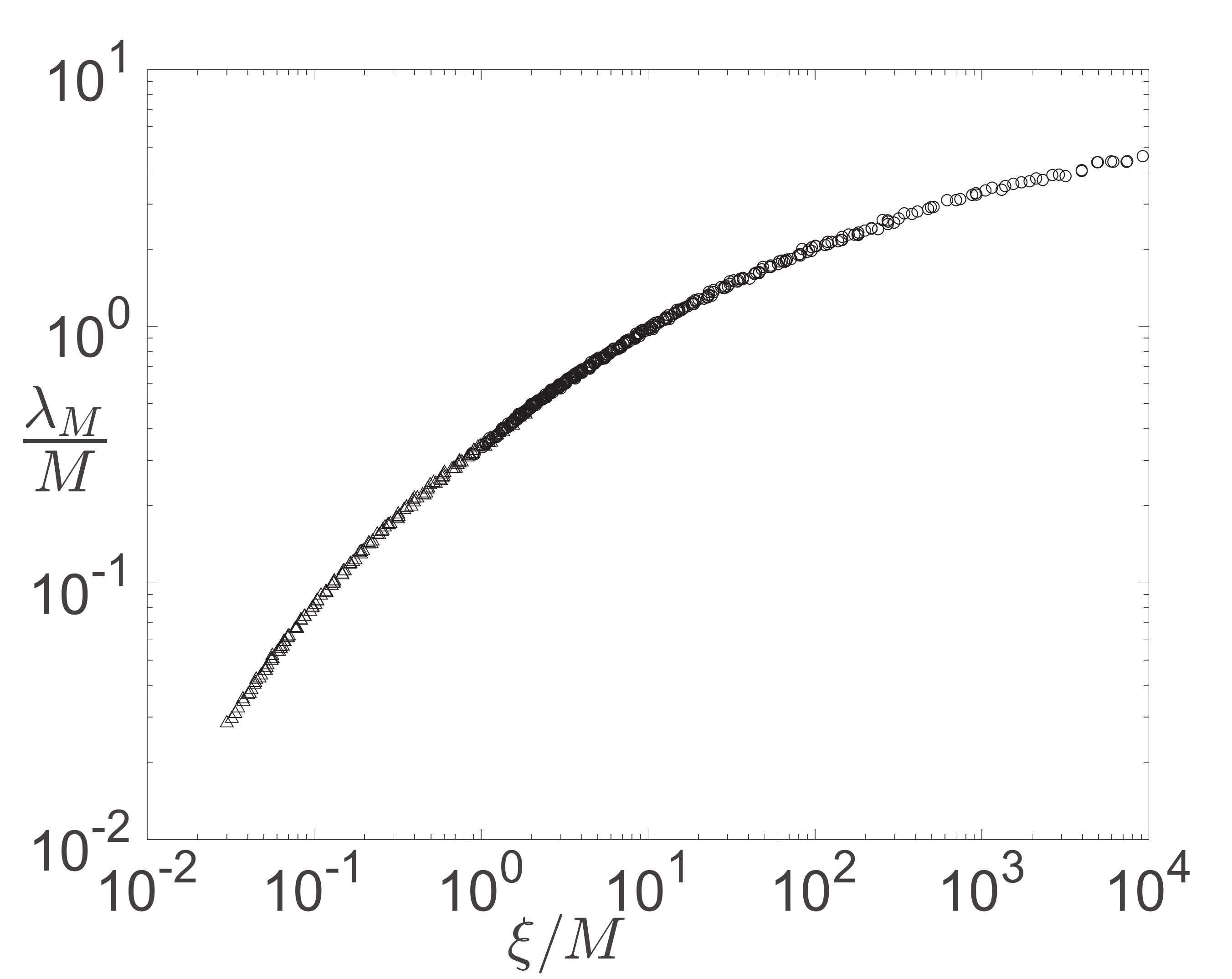}}
    \subfigure[]{
    \label{fig:localization3} 
    \includegraphics[width=0.32\textwidth]{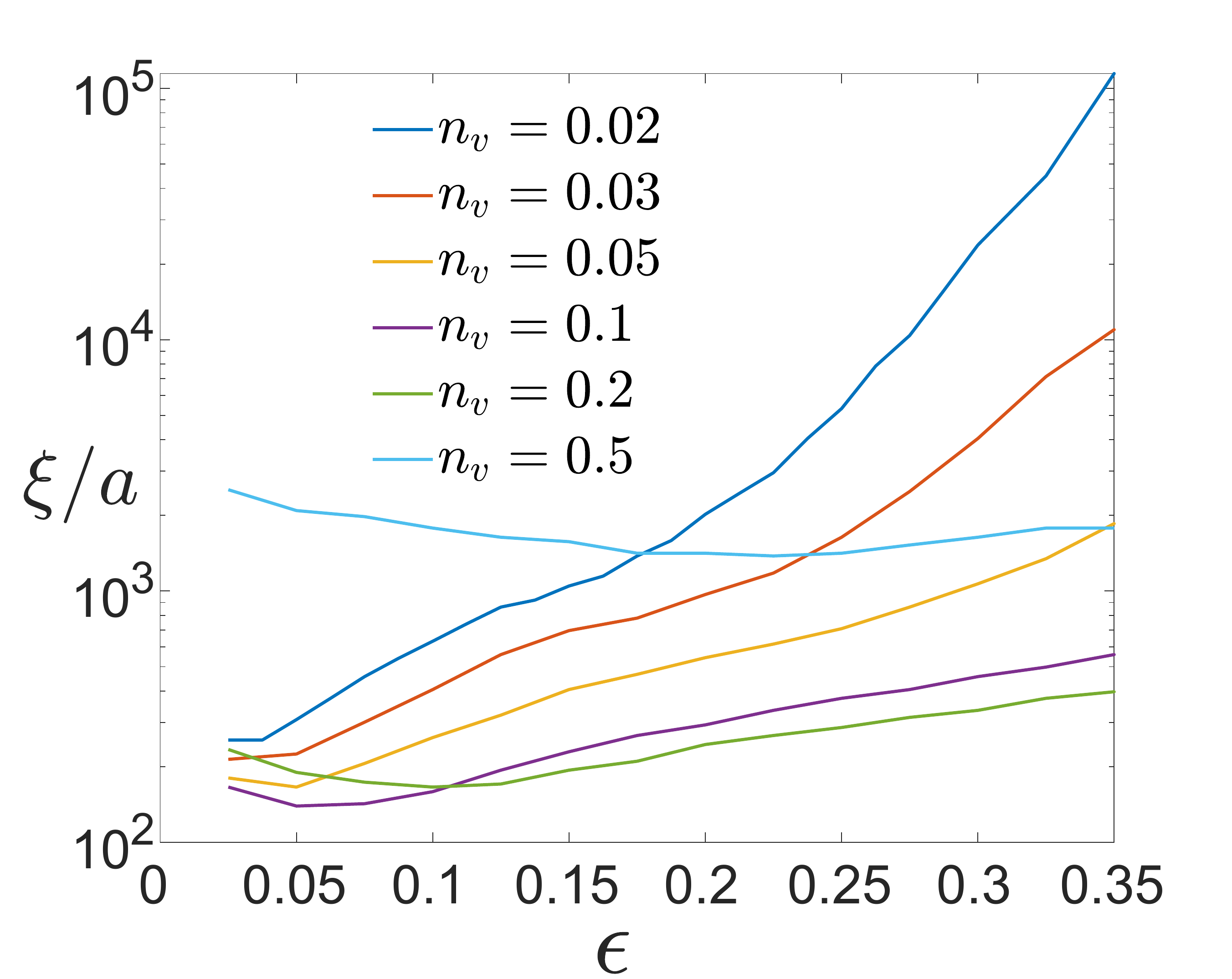}}
    \caption{(a) Log-log plot of $\lambda_M/M$ versus $M/a$ for $n_v=0.05$; we choose $M/(\sqrt{3}a)=48,60,72,90$ for $0.05\leq \epsilon\leq 0.5$ (black, the arrow denotes the ascending order and $\epsilon$ increases by 0.025 between adjacent lines), $M/(\sqrt{3}a)=48,49,50,60,61,62,72,73,74,90,91,92$ for $\epsilon=0.025$ (red) and $\epsilon=0$ (blue). (b) One-parameter scaling of the Mackinnon-Kramer parameter $\lambda_M/M$. The circles denote the data for $\epsilon\in[0.025,0.5] (n_v=0.02,0.03,0.05,0.1,0.2,0.5)$; to linearly fit the scaling function at small $\xi/M$ we also compute the parameter range $\epsilon \in [2.96,3] (n_v=0.05)$, represented by the triangles. (c) The 2-D localization length $\xi$ extracted by the scaling analysis as a function of $\epsilon$. }
    \label{fig:localization}
\end{figure*}
In graphene, the inter(intra)-valley scattering leads to (anti-)localization due to the constructive(destructive) quantum interference of backscattering amplitude \cite{Castro2009,DasSarma2011,Peres2010}. In RPFHM, the $\pi$-fluxes may be regarded as short-range scatters that mix different valleys and are hence expected to localize the Dirac fermions. According to the theory of weak-localization \cite{Lee1985}, the conductivity acquires a logarithmic quantum correction, namely
\begin{equation}
    \sigma_\text{G}=\sigma_\text{sc}-\frac{\alpha e^2}{h}\ln \frac{L}{l_0},
\end{equation}
where $l_0$ is the lower cutoff length of the diffusive transport regime and is comparable to the mean free path $l_e$. Our results in Fig. \ref{fig:Conductivity2} suggest that the conductivity decreases logarithmically with the system size, thus supporting the existence of weak localization. By numerical fitting, we obtain $\alpha\approx 0.39$, which is close to $\alpha=1/\pi$ predicted for the orthogonal universality class AI. 

\subsection{Anderson Localization}
To further study the effects of localization in the RPFHM, we use the transfer matrix method \cite{Mackinnon1981,MacKinnon1983} to compute the Lyapunov exponent of the system around its Dirac point. 
We consider the same geometry as shown in Fig. \ref{fig:setup}, in which the system can be divided into successive slices labeled by $n$. The Schr\"odinger equation at given energy $E$ can be written in the form of
\begin{equation}
    \left(\begin{array}{c}
         |\Psi_{n+1}\rangle\\
          |\Psi_n\rangle
    \end{array}\right)=T_n    \left(\begin{array}{c}
         |\Psi_{n}\rangle\\
          |\Psi_{n-1}\rangle
    \end{array}\right),
\end{equation}
where $|\Psi_n\rangle$ is the wavefunction of slice $n$, $T_n$ is the transfer matrix
\begin{equation}
    T_n=\left(\begin{array}{cc}
       H_{n,n+1}^{-1}(E-H_{n,n})  &  -H_{n,n+1}^{-1} H_{n,n-1} \\
        \mathds{1} & 0
    \end{array}\right),
\end{equation}
and $H_{m,n}$ is the Hamiltonian matrix between slice $m$ and $n$. By iteration, one obtains
\begin{equation}
        \left(\begin{array}{c}
         |\Psi_{n+1}\rangle\\
          |\Psi_n\rangle
    \end{array}\right)=M_n    \left(\begin{array}{c}
         |\Psi_{1}\rangle\\
          |\Psi_{0}\rangle
    \end{array}\right),
\end{equation}
where $M_n=T_nT_{n-1}\cdots T_2T_1$. There exists a limiting matrix $M_\infty=\text{lim}_{n\rightarrow\infty}(M_nM_n^\dagger)^{1/(2n)}$, which has eigenvalues $e^{\gamma_i}$, where $\gamma_i$ is the Lyapunov exponent. The Lyapunov exponents must come in opposite pairs and the quasi-one-dimensional localization length $\lambda_M$ can be defined as the inverse of the smallest positive Lyapunov exponent.
In all the simulations, we implement Gram-Schmidt orthonormalization every 8 steps for numerical stability (e.g. see \cite{Tadjine2018} and references herein) and the relative error of all the data is controlled within $\epsilon_\text{error}\lesssim 1\%$.

According to the one-parameter scaling theory of Anderson localization \cite{Mackinnon1981,MacKinnon1983}, the MacKinnon-Kramer parameter $\lambda_M/M$ is a single-parameter function of $\xi/M$, namely $\lambda_M/M=f(\xi/M)$, where the 2-D localization length $\xi$ depends on all parameters except $M$, which are $n_v$ and $\epsilon$ in our case. For the system in the orthogonal universality class AI, the scaling function $f(x)$ is a monotonically increasing function and $f(x)\rightarrow x$ when $x\rightarrow0$. We find that all states near the Dirac point except $\epsilon=0$ are consistent with this hypothesis (see Fig. \ref{fig:localization}), confirming the disorder-free localization also observed earlier. The extracted 2-D localization length $\xi$ shown in Fig. \ref{fig:localization3} indicated that the low-energy states could be strongly localized in mesoscopic systems with size larger than $10^2-10^3a$. 

The case of the exact Dirac point $\epsilon=0$ requires special attention, as the RPFHM belongs to the chiral orthogonal class BDI at this special energy. As shown in Fig. \ref{fig:localization1}, we observe significant oscillations of $\lambda_M/M$ at $\epsilon=0$ (blue line) due to the finite-size effects, in contrast to the $\epsilon=0.025$ (red line) case where the fluctuations are negligible. The oscillation has a period of three, which is likely related to the quasi-periodicity of the finite-size gap in a clean armchair-type graphene nanoribbon \cite{Nakada1996,Peres2006_2}. Similar behaviors have also been seen in random flux models on the square lattice, where $\lambda_M/M$ is sensitive to the parity of the width (in units of lattice spacing) \cite{Markos2007,Furusaki1999,Schweitzer2008,Tadjine2018}. It seems that the amplitude of oscillations tends to zero and $\lambda_M/M$ remains finite when $M\rightarrow\infty$, supporting the existence of critical and delocalized state at the band center of chiral metals \cite{Gade1991,Gade1993,Hatsugai1997,Furusaki1999,Ferreira2015}. A definite conclusion requires careful numerical analysis with a substantially larger system size and fine energy resolution, which is beyond the scope of this work. We note that this band-centered state may hardly affect the transport at realistic temperatures as it only exists within a very narrow energy window, making its identification challenging.

\section{Thermal conductivity of the Kitaev model}
The low-temperature behavior of thermal conductivity $\kappa_\text{K}$ of the Kitaev model can be inferred using earlier results. We first briefly introduce the Kitaev model \cite{Kitaev2006}
\begin{equation}
    H_\text{K}=\sum_{\langle ij\rangle_\lambda}J_\lambda  \sigma_i^\lambda\sigma_j^\lambda, \label{Kitaev}
\end{equation}
where Pauli matrices $\vec{\sigma_i}=(\sigma^x_i,\sigma^y_i,\sigma^z_i)$ describe the spin degrees of freedom on site $i$,  $\langle ij\rangle_\lambda$ denotes the $\lambda$-type nearest-neighbor bond ($\lambda=x,y,z$) between site $i$ and $j$ and each bond is only summed once, as shown in Fig. \ref{fig:Kitaev}. In this work, we are only interested in the isotropic case so we set $J=J_x=J_y=J_z$ hereafter. The Kitaev model can be exactly solved by mapping spins to Majorana fermions and the Hamiltonian becomes 
\begin{equation}
    H_\text{KSL}=\sum_{\langle i,j\rangle} \frac{iJ\tilde{u}_{ij}}{2}\gamma_i \gamma_j,\label{KitaevMajorana}
\end{equation}
where the Majorana fermions satisfy anticommutation relations $\{\gamma_i,\gamma_j\}=2\delta_{ij}$, and the bond operator $\tilde{u}_{ij}=-\tilde{u}_{ji}=\pm 1$ acts as a $Z_2$ gauge field. 

After making a gauge transformation $f_j^\dagger\rightarrow -if_j^\dagger, j\in A$ sublattice, followed by a Majorana transformation $f_{i}^\dagger=(\gamma_i^1+i \gamma_i^2)/2$ to Eq. (\ref{RPFHM}), one can see the RPFHM (\ref{RPFHM}) is exactly two copies of the Kitaev model (\ref{KitaevMajorana}), with $t=2J$. Therefore the thermal conductivity of the Kitaev model (\ref{Kitaev}) is half of that of the RPFHM (\ref{RPFHM}), namely $\kappa_\text{G}=2\kappa_\text{K}$, provided both have the same flux configuration. At thermal equilibrium, $\kappa_{\text{K}}$ can be expressed as the weighted average of thermal conductivity $ \kappa_{\text{K}}^{\tinyvarhexagon}$ over every flux configuration $\varhexagon$, namely $\kappa_{\text{K}}=(\sum_{\tinyvarhexagon} Z_{\tinyvarhexagon}\kappa_{\text{K}}^{\tinyvarhexagon})/(\sum_{\tinyvarhexagon} Z_{\tinyvarhexagon})$. Here $Z_{\tinyvarhexagon}=\text{Tr}e^{-\beta H_{\tinyvarhexagon}}$ is the partition function for specific flux configuration $\varhexagon$ described by the Hamiltonian $H_{\tinyvarhexagon}$, and $\sum_{\tinyvarhexagon}$ denotes sum over all flux configurations. In the Kitaev model the ground state contains no flux \cite{Lieb1994}, while at temperatures much lower than the vison gap $\Delta_v\approx 0.1536 J$ \cite{Kitaev2006}, thermally excited visons are dilute and may be regarded as uncorrelated on different plaquettes, with flux density approximately given by
\begin{equation}
    n_v=\frac{1}{e^{ \Delta_v/k_BT}+1}. \label{nv}
\end{equation}
Here we emphasize that although the calculation of $\kappa_{\text{K}}$ seems a quenched disorder problem, it is an artifact as the bond variables $u_{ij}$ commute with the Hamiltonian. In fact, it is still an annealed average problem, as the free energy is given by $\mathcal{F}=-k_B T \ln \sum_{\tinyvarhexagon} Z_{\tinyvarhexagon}$, where the logarithm is taken after the sum over all flux configurations is computed. 

The thermal conductivity $\kappa_\text{G}$ of RPFHM is linked to the electrical conductivity $\sigma_G$ by the Wiedemann–Franz (WF) law $\kappa_\text{G}=\sigma_\text{G}\mathcal{L}T$ where $\mathcal{L}$ is the Lorentz number. For a non-interacting Fermi liquid, the Lorentz number is a universal constant $\mathcal{L}_0=\pi^2k_B^2/3e^2$, while around the Dirac point, the Lorentz number may be a few times larger than $\mathcal{L}_0$ even for the non-interacting case \cite{Saito2007,Rycerz2021}. Combining the above arguments, one obtains
\begin{equation}
    \frac{\kappa_\text{K}\left(T\right)}{\sigma_\text{G}(n_v,T,\mu=0)}= \frac{\mathcal{L}}{2}T,\label{KThermalRelation}
\end{equation}
where $\sigma_\text{G}(n_v,T,\mu)$ denotes the DC conductivity of RPFHM, which has flux density $n_v$, Fermi level $\mu$ and temperature $T$.

We shall now demonstrate the low-temperature behavior of $\kappa_\text{K}$ using Eq. (\ref{KThermalRelation}). As $k_B T\gg|\mu|$, the conduction is mainly contributed by the thermally excited quasiparticles with energy $E\sim k_B T$. When $k_B T\ll \Delta_v$, we observe that the thermal de Broglie wavelength $\lambda_\text{th}\sim T^{-1}$ is much shorter than the mean flux distance $r_m\sim e^{\Delta_v/2k_BT}$. This indicates that the transport is away from the Dirac point and in the semiclassical regime. As the localization length is exponentially large $\xi\approx l_0e^{\pi k_F l_e}$ in this regime, the localization effects are negligible for realistic system size, or when the phase coherence length is much smaller than the localization length $L_\phi\ll \xi$ due to the coupling to environments. Substituting the thermal de Broglie wavevector $k_\text{th}=2\pi/\lambda_\text{th}$ to Eq. (\ref{DCconductivity2}) (see appendix \ref{sec:appendix} for the justification), which is applicable when $k_\text{th}r_m\gg 1$, and using Eqs. (\ref{nv})(\ref{KThermalRelation}), one obtains diverging thermal conductivity $\kappa_\text{K}\sim T^3 e^{\Delta_v/k_B T}$ at low temperatures. When $k_B T\sim \Delta_v\approx 0.15J$, which corresponds to $\lambda_\text{th}\gtrsim r_m$, the thermal transport is in the quantum regime and localization effects are non-negligible. We assume both the system size $L$ and phase coherence length $L_\phi$ is much larger than the thermal localization length $\xi_T=\xi(n_v(T),k_\text{th}(T))$, which is estimated to be a few hundred of lattice spacing from Fig. \ref{fig:localization3}. In this case, the Majorana fermions are strongly localized and significantly suppressed thermal conductivity $\kappa_\text{K}/T\ll k_B^2/2\pi\hbar$ may be observable. This is in contrast with the low-temperature case $k_B T\ll\Delta_v$ where $\kappa_\text{K}/T\gg k_B^2/2\pi\hbar$ and analogous to the large resistivity $\rho\gg h/e^2$ observed in graphene due to the strong localization \cite{Ponomarenko2011,Moser2010,Yanik2021}. We note that it is difficult to obtain $\kappa_\text{K}$ at higher temperature $k_B T\sim J \gg \Delta_v$ using our calculations, as the thermal transport is also contributed by quasiparticles at high energies that are not considered in this work. Besides, in this regime visons can no longer be regarded as dilute, and the validity of Eq. (\ref{nv}) becomes questionable.

\section{Conclusion}
In this work, we have utilized a combination of numerical methods to investigate the transport of RPFHM near the Dirac point. Our results reveal that when the wavelength of Dirac fermion is much shorter than the average flux spacing, the semiclassical DC conductivity is quadratic in the Fermi momentum and inversely proportional to the flux density. We have also demonstrated that intervalley scattering by the $\pi$-fluxes leads to weak (strong) localization of Dirac fermions at high (low) Fermi energy. 

\begin{figure}
    \centering
    \includegraphics[width= 0.48 \textwidth]{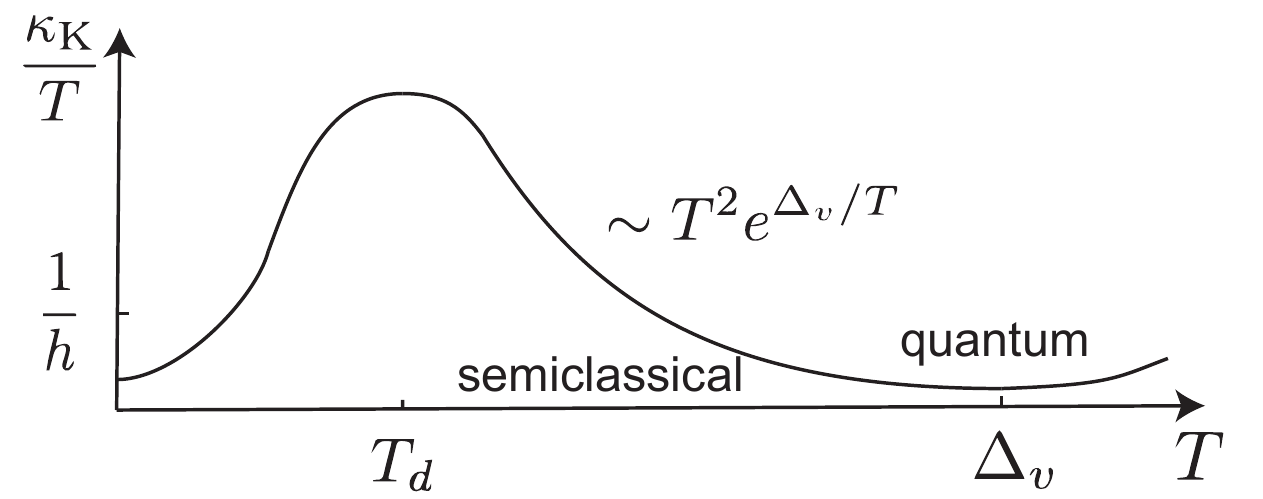}
    \caption{Schematics of the low-temperature thermal conductivity versus temperature. Note that the Boltzmann constant $k_B=1$.}
    \label{fig:kappavsT}
\end{figure}
Our results imply that the thermal transport of the Kitaev model is semiclassical at low temperatures $k_B T\ll \Delta_v$, and that the thermal conductivity diverges as $\kappa_\text{K}\sim T^3 e^{\Delta_v/k_B T}$. When $k_B T\sim \Delta_v$, the itinerant Majorana fermions may be strongly localized, leading to a significantly suppressed thermal conductivity $\kappa_\text{K}/T\ll k_B^2/2\pi\hbar$. Our predictions are expected to hold even in the presence of disorder as long as $\tau_v^{-1}\gg\tau_d^{-1}$, where $\tau_v^{-1}(\tau_d^{-1})$ is the scattering rate due to vison (disorder). When the temperature is much smaller than the crossover temperature $T_d$ at which $\tau_d^{-1}=\tau_v^{-1}$, the disorder becomes the dominant source of elastic scattering and the transport is similar to that of an undoped graphene. Consequently, $\kappa_\text{K}/T$ should reduce with decreasing temperature and finally saturates to a value comparable or smaller to $1/2\pi\hbar$ \cite{Castro2009,DasSarma2011,Peres2010}. Such temperature dependence, as shown in Fig. \ref{fig:kappavsT}, was unseen in previous numerical simulations and may be observable in a clean sample at very low temperatures. In realistic Kitaev materials, non-Kitaev-type interactions also allow the hopping of visons and may have important consequences \cite{Joy2022,Chen2023}, which are left for future studies.

\acknowledgements
I would like to thank P. Coleman for giving valuable feedback on the manuscript and J. Nasu for fruitful discussions.  The author acknowledges the support of two grant funding agencies, which funded two distinct components of the research carried out at different times.  The early calculation of the conductivity of the random $\pi$-flux honeycomb lattice was supported by grant DE-SC0020353 funded by the U.S. Department of Energy, Office of Science. A  later component of the research, calculating the two-dimensional localization length from transfer matrices, was supported by the NSF Division of Materials Research grant NSF grant DMR-1830707. 

\appendix

\section{The DC conductivity of RPFHM at intermediate temperatures}\label{sec:appendix}
\begin{figure}[b]
    \centering
    \includegraphics[width= 0.48\textwidth]{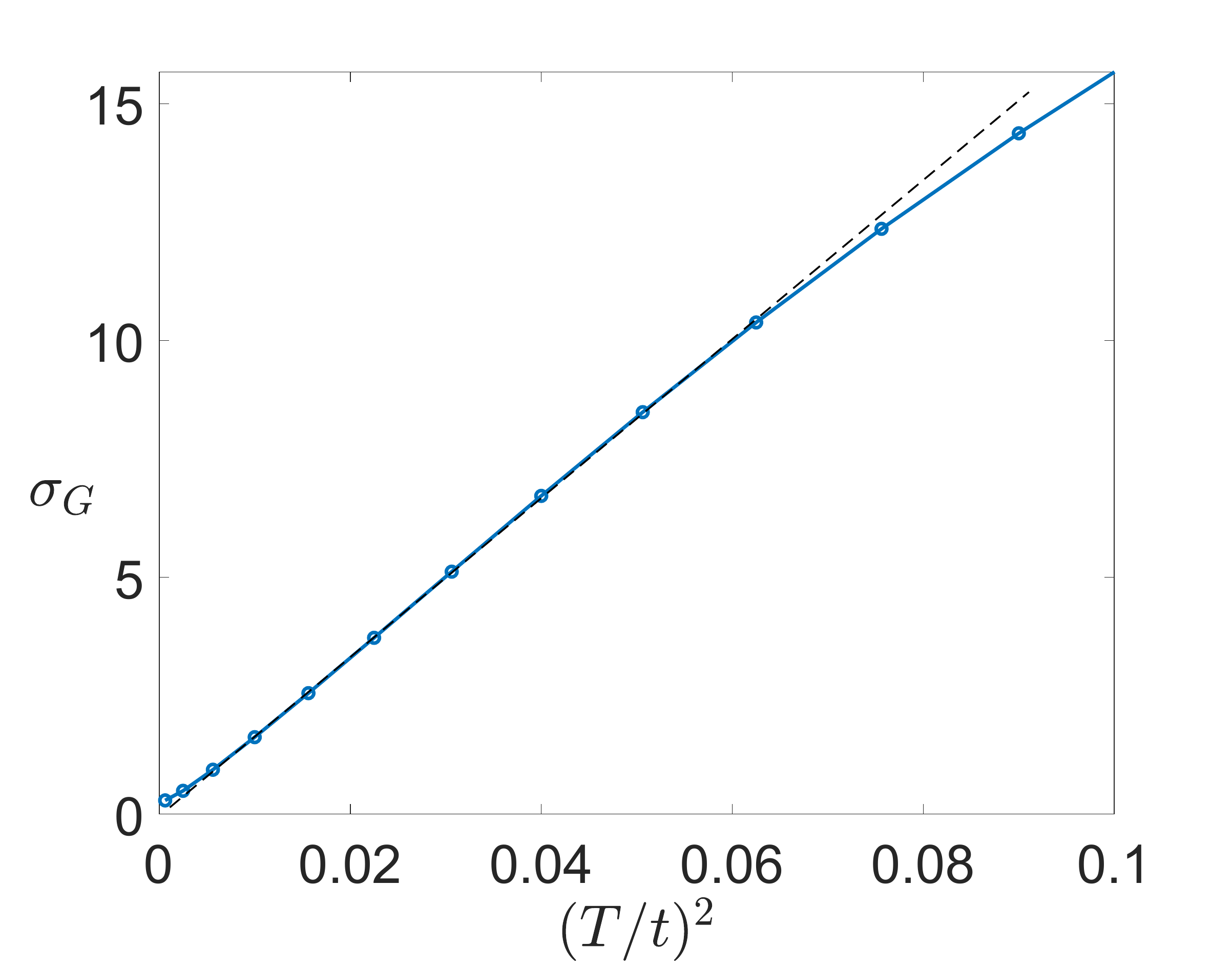}
    \caption{The DC conductivity $\sigma_\text{G}$ (in units of $e^2/h$) of the RPFHM versus $(T/t)^2$ for $n_v=0.01$ when $L\approx M=108\sqrt{3}a$ (the Boltzmann constant $k_B$ is set to unity).}
    \label{fig:GvsT}
\end{figure}
In this appendix, we calculate the finite-temperature DC conductivity of the RPFHM at half-filling and show that it is proportional to $T^2$, when $a/r_m \lesssim k_B T/t\ll 1$. This justifies that, in this temperature regime, the finite-temperature conductivity of RPFHM at half-filling can be inferred from the zero-temperature conductivity of RPFHM at finite Fermi energy, by replacing the Fermi momentum $k_F$ in Eq. (\ref{DCconductivity2}) with the thermal de Broglie wavevector $k_\text{th}=2\pi/\lambda_\text{th}$. \\
The current can be calculated using the Landauer formula
\begin{equation}
    I=\frac{e}{h}\int dE \left[f(E-\mu_L)-f(E-\mu_R)\right] \mathbf{T}(E),
\end{equation}
where $f(E)=1/(e^{E/k_B T}+1)$ is the Fermi distribution function, and $\mu_L(\mu_R)$ is the chemical potential of the left(right) lead. In the linear response regime, the above equation gives the conductance
\begin{equation}
    G=\frac{e^2}{h} \int dE \mathbf{T}(E)\left(-\frac{df(E)}{dE}\right),
\end{equation}
where we have used $\mu_L+\mu_R=0$ and $\mu_L-\mu_R=eV$. The finite-temperature conductivity can then be extracted using Eq. (\ref{DCconductivity}) for fixed flux density $n_v$ and system size, and the result is shown in Fig. \ref{fig:GvsT}. We note that $\sigma_G$ shows a $T^2$ temperature dependence at intermediate temperatures $a/r_m \lesssim k_B T/t\ll 1$, which originates from the $k_F^2$ dependence of the zero-temperature conductivity (see Eq. (\ref{DCconductivity2})). The deviation from the $T^2$ dependence at low temperatures $k_B T/t\lesssim a/r_m$ is due to the quantum transport near the Dirac point, for which Eq. (\ref{DCconductivity2}) no longer applies. The discrepancy at high temperatures $k_B T/t \sim 1$ is attributed to the nonlinearity of Dirac dispersion and the thermally excited carriers above the Van Hove point. We remark that although the Landauer approach fails near the Dirac point $E=0$, it does not affect our finite-temperature calculation significantly, as long as $a/r_m \lesssim k_B T/t$.

\bibliography{Reference}

\begin{thebibliography}{67}%
\makeatletter
\providecommand \@ifxundefined [1]{%
 \@ifx{#1\undefined}
}%
\providecommand \@ifnum [1]{%
 \ifnum #1\expandafter \@firstoftwo
 \else \expandafter \@secondoftwo
 \fi
}%
\providecommand \@ifx [1]{%
 \ifx #1\expandafter \@firstoftwo
 \else \expandafter \@secondoftwo
 \fi
}%
\providecommand \natexlab [1]{#1}%
\providecommand \enquote  [1]{``#1''}%
\providecommand \bibnamefont  [1]{#1}%
\providecommand \bibfnamefont [1]{#1}%
\providecommand \citenamefont [1]{#1}%
\providecommand \href@noop [0]{\@secondoftwo}%
\providecommand \href [0]{\begingroup \@sanitize@url \@href}%
\providecommand \@href[1]{\@@startlink{#1}\@@href}%
\providecommand \@@href[1]{\endgroup#1\@@endlink}%
\providecommand \@sanitize@url [0]{\catcode `\\12\catcode `\$12\catcode
  `\&12\catcode `\#12\catcode `\^12\catcode `\_12\catcode `\%12\relax}%
\providecommand \@@startlink[1]{}%
\providecommand \@@endlink[0]{}%
\providecommand \url  [0]{\begingroup\@sanitize@url \@url }%
\providecommand \@url [1]{\endgroup\@href {#1}{\urlprefix }}%
\providecommand \urlprefix  [0]{URL }%
\providecommand \Eprint [0]{\href }%
\providecommand \doibase [0]{https://doi.org/}%
\providecommand \selectlanguage [0]{\@gobble}%
\providecommand \bibinfo  [0]{\@secondoftwo}%
\providecommand \bibfield  [0]{\@secondoftwo}%
\providecommand \translation [1]{[#1]}%
\providecommand \BibitemOpen [0]{}%
\providecommand \bibitemStop [0]{}%
\providecommand \bibitemNoStop [0]{.\EOS\space}%
\providecommand \EOS [0]{\spacefactor3000\relax}%
\providecommand \BibitemShut  [1]{\csname bibitem#1\endcsname}%
\let\auto@bib@innerbib\@empty
\bibitem [{\citenamefont {Anderson}(1973)}]{Anderson1973}%
  \BibitemOpen
  \bibfield  {author} {\bibinfo {author} {\bibfnamefont {P.~W.}\ \bibnamefont
  {Anderson}},\ }\bibfield  {title} {\bibinfo {title} {Resonating valence
  bonds: A new kind of insulator?},\ }\href
  {https://doi.org/10.1016/0025-5408(73)90167-0} {\bibfield  {journal}
  {\bibinfo  {journal} {Materials Research Bulletin}\ }\textbf {\bibinfo
  {volume} {8}},\ \bibinfo {pages} {153} (\bibinfo {year} {1973})}\BibitemShut
  {NoStop}%
\bibitem [{\citenamefont {Savary}\ and\ \citenamefont
  {Balents}(2016)}]{Savary2016}%
  \BibitemOpen
  \bibfield  {author} {\bibinfo {author} {\bibfnamefont {L.}~\bibnamefont
  {Savary}}\ and\ \bibinfo {author} {\bibfnamefont {L.}~\bibnamefont
  {Balents}},\ }\bibfield  {title} {\bibinfo {title} {Quantum spin liquids: a
  review},\ }\href {https://doi.org/10.1088/0034-4885/80/1/016502} {\bibfield
  {journal} {\bibinfo  {journal} {Reports on Progress in Physics}\ }\textbf
  {\bibinfo {volume} {80}},\ \bibinfo {pages} {016502} (\bibinfo {year}
  {2016})}\BibitemShut {NoStop}%
\bibitem [{\citenamefont {Zhou}\ \emph {et~al.}(2017)\citenamefont {Zhou},
  \citenamefont {Kanoda},\ and\ \citenamefont {Ng}}]{Zhou2017}%
  \BibitemOpen
  \bibfield  {author} {\bibinfo {author} {\bibfnamefont {Y.}~\bibnamefont
  {Zhou}}, \bibinfo {author} {\bibfnamefont {K.}~\bibnamefont {Kanoda}},\ and\
  \bibinfo {author} {\bibfnamefont {T.-K.}\ \bibnamefont {Ng}},\ }\bibfield
  {title} {\bibinfo {title} {Quantum spin liquid states},\ }\href
  {https://doi.org/10.1103/RevModPhys.89.025003} {\bibfield  {journal}
  {\bibinfo  {journal} {Rev. Mod. Phys.}\ }\textbf {\bibinfo {volume} {89}},\
  \bibinfo {pages} {025003} (\bibinfo {year} {2017})}\BibitemShut {NoStop}%
\bibitem [{\citenamefont {lieb}(2006)}]{Kitaev2006}%
  \BibitemOpen
  \bibfield  {author} {\bibinfo {author} {\bibfnamefont {A.}~\bibnamefont
  {lieb}},\ }\bibfield  {title} {\bibinfo {title} {Anyons in an exactly solved
  model and beyond},\ }\href
  {https://doi.org/https://doi.org/10.1016/j.aop.2005.10.005} {\bibfield
  {journal} {\bibinfo  {journal} {Annals of Physics}\ }\textbf {\bibinfo
  {volume} {321}},\ \bibinfo {pages} {2 } (\bibinfo {year} {2006})}\BibitemShut
  {NoStop}%
\bibitem [{\citenamefont {Jackeli}\ and\ \citenamefont
  {Khaliullin}(2009)}]{Jackeli2009}%
  \BibitemOpen
  \bibfield  {author} {\bibinfo {author} {\bibfnamefont {G.}~\bibnamefont
  {Jackeli}}\ and\ \bibinfo {author} {\bibfnamefont {G.}~\bibnamefont
  {Khaliullin}},\ }\bibfield  {title} {\bibinfo {title} {{Mott} insulators in
  the strong spin-orbit coupling limit: {From} {Heisenberg} to a quantum
  compass and {Kitaev} models},\ }\href
  {https://doi.org/10.1103/PhysRevLett.102.017205} {\bibfield  {journal}
  {\bibinfo  {journal} {Phys. Rev. Lett.}\ }\textbf {\bibinfo {volume} {102}},\
  \bibinfo {pages} {017205} (\bibinfo {year} {2009})}\BibitemShut {NoStop}%
\bibitem [{\citenamefont {Winter}\ \emph {et~al.}(2017)\citenamefont {Winter},
  \citenamefont {Tsirlin}, \citenamefont {Daghofer}, \citenamefont {van~den
  Brink}, \citenamefont {Singh}, \citenamefont {Gegenwart},\ and\ \citenamefont
  {Valent{\'\i}}}]{Winter2017}%
  \BibitemOpen
  \bibfield  {author} {\bibinfo {author} {\bibfnamefont {S.~M.}\ \bibnamefont
  {Winter}}, \bibinfo {author} {\bibfnamefont {A.~A.}\ \bibnamefont {Tsirlin}},
  \bibinfo {author} {\bibfnamefont {M.}~\bibnamefont {Daghofer}}, \bibinfo
  {author} {\bibfnamefont {J.}~\bibnamefont {van~den Brink}}, \bibinfo {author}
  {\bibfnamefont {Y.}~\bibnamefont {Singh}}, \bibinfo {author} {\bibfnamefont
  {P.}~\bibnamefont {Gegenwart}},\ and\ \bibinfo {author} {\bibfnamefont
  {R.}~\bibnamefont {Valent{\'\i}}},\ }\bibfield  {title} {\bibinfo {title}
  {Models and materials for generalized kitaev magnetism},\ }\href
  {https://dx.doi.org/10.1088/1361-648X/aa8cf5} {\bibfield  {journal} {\bibinfo
   {journal} {Journal of Physics: Condensed Matter}\ }\textbf {\bibinfo
  {volume} {29}},\ \bibinfo {pages} {493002} (\bibinfo {year}
  {2017})}\BibitemShut {NoStop}%
\bibitem [{\citenamefont {Trebst}\ and\ \citenamefont
  {Hickey}(2022)}]{Trebst2022}%
  \BibitemOpen
  \bibfield  {author} {\bibinfo {author} {\bibfnamefont {S.}~\bibnamefont
  {Trebst}}\ and\ \bibinfo {author} {\bibfnamefont {C.}~\bibnamefont
  {Hickey}},\ }\bibfield  {title} {\bibinfo {title} {Kitaev materials},\ }\href
  {https://doi.org/10.1016/j.physrep.2021.11.003} {\bibfield  {journal}
  {\bibinfo  {journal} {Physics Reports}\ }\textbf {\bibinfo {volume} {950}},\
  \bibinfo {pages} {1} (\bibinfo {year} {2022})}\BibitemShut {NoStop}%
\bibitem [{\citenamefont {Kasahara}\ \emph
  {et~al.}(2018{\natexlab{a}})\citenamefont {Kasahara}, \citenamefont {Sugii},
  \citenamefont {Ohnishi}, \citenamefont {Shimozawa}, \citenamefont
  {Yamashita}, \citenamefont {Kurita}, \citenamefont {Tanaka}, \citenamefont
  {Nasu}, \citenamefont {Motome}, \citenamefont {Shibauchi},\ and\
  \citenamefont {Matsuda}}]{Kasahara2018}%
  \BibitemOpen
  \bibfield  {author} {\bibinfo {author} {\bibfnamefont {Y.}~\bibnamefont
  {Kasahara}}, \bibinfo {author} {\bibfnamefont {K.}~\bibnamefont {Sugii}},
  \bibinfo {author} {\bibfnamefont {T.}~\bibnamefont {Ohnishi}}, \bibinfo
  {author} {\bibfnamefont {M.}~\bibnamefont {Shimozawa}}, \bibinfo {author}
  {\bibfnamefont {M.}~\bibnamefont {Yamashita}}, \bibinfo {author}
  {\bibfnamefont {N.}~\bibnamefont {Kurita}}, \bibinfo {author} {\bibfnamefont
  {H.}~\bibnamefont {Tanaka}}, \bibinfo {author} {\bibfnamefont
  {J.}~\bibnamefont {Nasu}}, \bibinfo {author} {\bibfnamefont {Y.}~\bibnamefont
  {Motome}}, \bibinfo {author} {\bibfnamefont {T.}~\bibnamefont {Shibauchi}},\
  and\ \bibinfo {author} {\bibfnamefont {Y.}~\bibnamefont {Matsuda}},\
  }\bibfield  {title} {\bibinfo {title} {Unusual thermal hall effect in a
  {Kitaev} spin liquid candidate
  $\ensuremath{\alpha}\text{\ensuremath{-}}{\mathrm{rucl}}_{3}$},\ }\href
  {https://doi.org/10.1103/PhysRevLett.120.217205} {\bibfield  {journal}
  {\bibinfo  {journal} {Phys. Rev. Lett.}\ }\textbf {\bibinfo {volume} {120}},\
  \bibinfo {pages} {217205} (\bibinfo {year} {2018}{\natexlab{a}})}\BibitemShut
  {NoStop}%
\bibitem [{\citenamefont {Kasahara}\ \emph
  {et~al.}(2018{\natexlab{b}})\citenamefont {Kasahara}, \citenamefont
  {Ohnishi}, \citenamefont {Mizukami}, \citenamefont {Tanaka}, \citenamefont
  {Ma}, \citenamefont {Sugii}, \citenamefont {Kurita}, \citenamefont {Tanaka},
  \citenamefont {Nasu}, \citenamefont {Motome} \emph
  {et~al.}}]{Kasahara2018_2}%
  \BibitemOpen
  \bibfield  {author} {\bibinfo {author} {\bibfnamefont {Y.}~\bibnamefont
  {Kasahara}}, \bibinfo {author} {\bibfnamefont {T.}~\bibnamefont {Ohnishi}},
  \bibinfo {author} {\bibfnamefont {Y.}~\bibnamefont {Mizukami}}, \bibinfo
  {author} {\bibfnamefont {O.}~\bibnamefont {Tanaka}}, \bibinfo {author}
  {\bibfnamefont {S.}~\bibnamefont {Ma}}, \bibinfo {author} {\bibfnamefont
  {K.}~\bibnamefont {Sugii}}, \bibinfo {author} {\bibfnamefont
  {N.}~\bibnamefont {Kurita}}, \bibinfo {author} {\bibfnamefont
  {H.}~\bibnamefont {Tanaka}}, \bibinfo {author} {\bibfnamefont
  {J.}~\bibnamefont {Nasu}}, \bibinfo {author} {\bibfnamefont {Y.}~\bibnamefont
  {Motome}}, \emph {et~al.},\ }\bibfield  {title} {\bibinfo {title} {{Majorana
  quantization and half-integer thermal quantum Hall effect in a {Kitaev} spin
  liquid}},\ }\href {https://doi.org/10.1038/s41586-018-0274-0} {\bibfield
  {journal} {\bibinfo  {journal} {Nature}\ }\textbf {\bibinfo {volume} {559}},\
  \bibinfo {pages} {227} (\bibinfo {year} {2018}{\natexlab{b}})}\BibitemShut
  {NoStop}%
\bibitem [{\citenamefont {Yokoi}\ \emph {et~al.}(2021)\citenamefont {Yokoi},
  \citenamefont {Ma}, \citenamefont {Kasahara}, \citenamefont {Kasahara},
  \citenamefont {Shibauchi}, \citenamefont {Kurita}, \citenamefont {Tanaka},
  \citenamefont {Nasu}, \citenamefont {Motome}, \citenamefont {Hickey},
  \citenamefont {Trebst},\ and\ \citenamefont {Matsuda}}]{Yokoi2021}%
  \BibitemOpen
  \bibfield  {author} {\bibinfo {author} {\bibfnamefont {T.}~\bibnamefont
  {Yokoi}}, \bibinfo {author} {\bibfnamefont {S.}~\bibnamefont {Ma}}, \bibinfo
  {author} {\bibfnamefont {Y.}~\bibnamefont {Kasahara}}, \bibinfo {author}
  {\bibfnamefont {S.}~\bibnamefont {Kasahara}}, \bibinfo {author}
  {\bibfnamefont {T.}~\bibnamefont {Shibauchi}}, \bibinfo {author}
  {\bibfnamefont {N.}~\bibnamefont {Kurita}}, \bibinfo {author} {\bibfnamefont
  {H.}~\bibnamefont {Tanaka}}, \bibinfo {author} {\bibfnamefont
  {J.}~\bibnamefont {Nasu}}, \bibinfo {author} {\bibfnamefont {Y.}~\bibnamefont
  {Motome}}, \bibinfo {author} {\bibfnamefont {C.}~\bibnamefont {Hickey}},
  \bibinfo {author} {\bibfnamefont {S.}~\bibnamefont {Trebst}},\ and\ \bibinfo
  {author} {\bibfnamefont {Y.}~\bibnamefont {Matsuda}},\ }\bibfield  {title}
  {\bibinfo {title} {{Half-integer quantized anomalous thermal Hall effect in
  the Kitaev material candidate $\alpha$-RuCl$_3$}},\ }\href
  {https://doi.org/10.1126/science.aay5551} {\bibfield  {journal} {\bibinfo
  {journal} {Science}\ }\textbf {\bibinfo {volume} {373}},\ \bibinfo {pages}
  {568} (\bibinfo {year} {2021})}\BibitemShut {NoStop}%
\bibitem [{\citenamefont {Nasu}\ \emph {et~al.}(2017)\citenamefont {Nasu},
  \citenamefont {Yoshitake},\ and\ \citenamefont {Motome}}]{Nasu2017}%
  \BibitemOpen
  \bibfield  {author} {\bibinfo {author} {\bibfnamefont {J.}~\bibnamefont
  {Nasu}}, \bibinfo {author} {\bibfnamefont {J.}~\bibnamefont {Yoshitake}},\
  and\ \bibinfo {author} {\bibfnamefont {Y.}~\bibnamefont {Motome}},\
  }\bibfield  {title} {\bibinfo {title} {Thermal transport in the {Kitaev}
  model},\ }\href {https://doi.org/10.1103/PhysRevLett.119.127204} {\bibfield
  {journal} {\bibinfo  {journal} {Phys. Rev. Lett.}\ }\textbf {\bibinfo
  {volume} {119}},\ \bibinfo {pages} {127204} (\bibinfo {year}
  {2017})}\BibitemShut {NoStop}%
\bibitem [{\citenamefont {Metavitsiadis}\ \emph {et~al.}(2017)\citenamefont
  {Metavitsiadis}, \citenamefont {Pidatella},\ and\ \citenamefont
  {Brenig}}]{Metavitsiadis2017}%
  \BibitemOpen
  \bibfield  {author} {\bibinfo {author} {\bibfnamefont {A.}~\bibnamefont
  {Metavitsiadis}}, \bibinfo {author} {\bibfnamefont {A.}~\bibnamefont
  {Pidatella}},\ and\ \bibinfo {author} {\bibfnamefont {W.}~\bibnamefont
  {Brenig}},\ }\bibfield  {title} {\bibinfo {title} {Thermal transport in a
  two-dimensional {$Z_{2}$} spin liquid},\ }\href
  {https://doi.org/10.1103/PhysRevB.96.205121} {\bibfield  {journal} {\bibinfo
  {journal} {Phys. Rev. B}\ }\textbf {\bibinfo {volume} {96}},\ \bibinfo
  {pages} {205121} (\bibinfo {year} {2017})}\BibitemShut {NoStop}%
\bibitem [{\citenamefont {Pidatella}\ \emph {et~al.}(2019)\citenamefont
  {Pidatella}, \citenamefont {Metavitsiadis},\ and\ \citenamefont
  {Brenig}}]{Pidatella2019}%
  \BibitemOpen
  \bibfield  {author} {\bibinfo {author} {\bibfnamefont {A.}~\bibnamefont
  {Pidatella}}, \bibinfo {author} {\bibfnamefont {A.}~\bibnamefont
  {Metavitsiadis}},\ and\ \bibinfo {author} {\bibfnamefont {W.}~\bibnamefont
  {Brenig}},\ }\bibfield  {title} {\bibinfo {title} {Heat transport in the
  anisotropic kitaev spin liquid},\ }\href
  {https://doi.org/10.1103/PhysRevB.99.075141} {\bibfield  {journal} {\bibinfo
  {journal} {Phys. Rev. B}\ }\textbf {\bibinfo {volume} {99}},\ \bibinfo
  {pages} {075141} (\bibinfo {year} {2019})}\BibitemShut {NoStop}%
\bibitem [{\citenamefont {Kao}\ and\ \citenamefont {Perkins}(2021)}]{Kao2021}%
  \BibitemOpen
  \bibfield  {author} {\bibinfo {author} {\bibfnamefont {W.-H.}\ \bibnamefont
  {Kao}}\ and\ \bibinfo {author} {\bibfnamefont {N.~B.}\ \bibnamefont
  {Perkins}},\ }\bibfield  {title} {\bibinfo {title} {Disorder upon disorder:
  Localization effects in the {Kitaev} spin liquid},\ }\href
  {https://doi.org/https://doi.org/10.1016/j.aop.2021.168506} {\bibfield
  {journal} {\bibinfo  {journal} {Annals of Physics}\ }\textbf {\bibinfo
  {volume} {435}},\ \bibinfo {pages} {168506} (\bibinfo {year} {2021})},\
  \bibinfo {note} {special issue on Philip W. Anderson}\BibitemShut {NoStop}%
\bibitem [{\citenamefont {Nasu}\ and\ \citenamefont {Motome}(2020)}]{Nasu2020}%
  \BibitemOpen
  \bibfield  {author} {\bibinfo {author} {\bibfnamefont {J.}~\bibnamefont
  {Nasu}}\ and\ \bibinfo {author} {\bibfnamefont {Y.}~\bibnamefont {Motome}},\
  }\bibfield  {title} {\bibinfo {title} {Thermodynamic and transport properties
  in disordered {Kitaev} models},\ }\href
  {https://doi.org/10.1103/PhysRevB.102.054437} {\bibfield  {journal} {\bibinfo
   {journal} {Phys. Rev. B}\ }\textbf {\bibinfo {volume} {102}},\ \bibinfo
  {pages} {054437} (\bibinfo {year} {2020})}\BibitemShut {NoStop}%
\bibitem [{\citenamefont {Joy}\ and\ \citenamefont {Rosch}(2022)}]{Joy2022}%
  \BibitemOpen
  \bibfield  {author} {\bibinfo {author} {\bibfnamefont {A.~P.}\ \bibnamefont
  {Joy}}\ and\ \bibinfo {author} {\bibfnamefont {A.}~\bibnamefont {Rosch}},\
  }\bibfield  {title} {\bibinfo {title} {{Dynamics of Visons and Thermal Hall
  Effect in Perturbed Kitaev Models}},\ }\href
  {https://doi.org/10.1103/PhysRevX.12.041004} {\bibfield  {journal} {\bibinfo
  {journal} {Phys. Rev. X}\ }\textbf {\bibinfo {volume} {12}},\ \bibinfo
  {pages} {041004} (\bibinfo {year} {2022})}\BibitemShut {NoStop}%
\bibitem [{\citenamefont {Zheng}\ and\ \citenamefont
  {Brataas}(2021)}]{Zheng2021}%
  \BibitemOpen
  \bibfield  {author} {\bibinfo {author} {\bibfnamefont {J.-H.}\ \bibnamefont
  {Zheng}}\ and\ \bibinfo {author} {\bibfnamefont {A.}~\bibnamefont
  {Brataas}},\ }\bibfield  {title} {\bibinfo {title} {Controlling the {RKKY}
  interaction and heat transport in a {Kitaev} spin liquid via {$Z_2$} flux
  walls},\ }\href {https://doi.org/10.1103/PhysRevB.104.064437} {\bibfield
  {journal} {\bibinfo  {journal} {Phys. Rev. B}\ }\textbf {\bibinfo {volume}
  {104}},\ \bibinfo {pages} {064437} (\bibinfo {year} {2021})}\BibitemShut
  {NoStop}%
\bibitem [{\citenamefont {Chen}\ and\ \citenamefont
  {Villadiego}(2023)}]{Chen2023}%
  \BibitemOpen
  \bibfield  {author} {\bibinfo {author} {\bibfnamefont {C.}~\bibnamefont
  {Chen}}\ and\ \bibinfo {author} {\bibfnamefont {I.~S.}\ \bibnamefont
  {Villadiego}},\ }\bibfield  {title} {\bibinfo {title} {Nature of visons in
  the perturbed ferromagnetic and antiferromagnetic kitaev honeycomb models},\
  }\href {https://doi.org/10.1103/PhysRevB.107.045114} {\bibfield  {journal}
  {\bibinfo  {journal} {Phys. Rev. B}\ }\textbf {\bibinfo {volume} {107}},\
  \bibinfo {pages} {045114} (\bibinfo {year} {2023})}\BibitemShut {NoStop}%
\bibitem [{\citenamefont {Koyama}\ and\ \citenamefont
  {Nasu}(2021)}]{Koyama2021}%
  \BibitemOpen
  \bibfield  {author} {\bibinfo {author} {\bibfnamefont {S.}~\bibnamefont
  {Koyama}}\ and\ \bibinfo {author} {\bibfnamefont {J.}~\bibnamefont {Nasu}},\
  }\bibfield  {title} {\bibinfo {title} {Field-angle dependence of thermal hall
  conductivity in a magnetically ordered kitaev-heisenberg system},\ }\href
  {https://doi.org/10.1103/PhysRevB.104.075121} {\bibfield  {journal} {\bibinfo
   {journal} {Phys. Rev. B}\ }\textbf {\bibinfo {volume} {104}},\ \bibinfo
  {pages} {075121} (\bibinfo {year} {2021})}\BibitemShut {NoStop}%
\bibitem [{\citenamefont {Zhang}\ \emph {et~al.}(2021)\citenamefont {Zhang},
  \citenamefont {Chern},\ and\ \citenamefont {Kim}}]{Zhang2021}%
  \BibitemOpen
  \bibfield  {author} {\bibinfo {author} {\bibfnamefont {E.~Z.}\ \bibnamefont
  {Zhang}}, \bibinfo {author} {\bibfnamefont {L.~E.}\ \bibnamefont {Chern}},\
  and\ \bibinfo {author} {\bibfnamefont {Y.~B.}\ \bibnamefont {Kim}},\
  }\bibfield  {title} {\bibinfo {title} {Topological magnons for thermal hall
  transport in frustrated magnets with bond-dependent interactions},\ }\href
  {https://doi.org/10.1103/PhysRevB.103.174402} {\bibfield  {journal} {\bibinfo
   {journal} {Phys. Rev. B}\ }\textbf {\bibinfo {volume} {103}},\ \bibinfo
  {pages} {174402} (\bibinfo {year} {2021})}\BibitemShut {NoStop}%
\bibitem [{\citenamefont {Cookmeyer}\ and\ \citenamefont
  {Moore}(2018)}]{Cookmeyer2018}%
  \BibitemOpen
  \bibfield  {author} {\bibinfo {author} {\bibfnamefont {T.}~\bibnamefont
  {Cookmeyer}}\ and\ \bibinfo {author} {\bibfnamefont {J.~E.}\ \bibnamefont
  {Moore}},\ }\bibfield  {title} {\bibinfo {title} {{Spin-wave analysis of the
  low-temperature thermal Hall effect in the candidate Kitaev spin liquid
  $\ensuremath{\alpha}\ensuremath{-}\text{RuCl}_{3}$}},\ }\href
  {https://doi.org/10.1103/PhysRevB.98.060412} {\bibfield  {journal} {\bibinfo
  {journal} {Phys. Rev. B}\ }\textbf {\bibinfo {volume} {98}},\ \bibinfo
  {pages} {060412} (\bibinfo {year} {2018})}\BibitemShut {NoStop}%
\bibitem [{\citenamefont {Fradkin}(1986)}]{Fradkin1986}%
  \BibitemOpen
  \bibfield  {author} {\bibinfo {author} {\bibfnamefont {E.}~\bibnamefont
  {Fradkin}},\ }\bibfield  {title} {\bibinfo {title} {Critical behavior of
  disordered degenerate semiconductors. ii. spectrum and transport properties
  in mean-field theory},\ }\href {https://doi.org/10.1103/PhysRevB.33.3263}
  {\bibfield  {journal} {\bibinfo  {journal} {Phys. Rev. B}\ }\textbf {\bibinfo
  {volume} {33}},\ \bibinfo {pages} {3263} (\bibinfo {year}
  {1986})}\BibitemShut {NoStop}%
\bibitem [{\citenamefont {Lee}(1993)}]{Lee1993}%
  \BibitemOpen
  \bibfield  {author} {\bibinfo {author} {\bibfnamefont {P.~A.}\ \bibnamefont
  {Lee}},\ }\bibfield  {title} {\bibinfo {title} {Localized states in a d-wave
  superconductor},\ }\href {https://doi.org/10.1103/PhysRevLett.71.1887}
  {\bibfield  {journal} {\bibinfo  {journal} {Phys. Rev. Lett.}\ }\textbf
  {\bibinfo {volume} {71}},\ \bibinfo {pages} {1887} (\bibinfo {year}
  {1993})}\BibitemShut {NoStop}%
\bibitem [{\citenamefont {Peres}\ \emph
  {et~al.}(2006{\natexlab{a}})\citenamefont {Peres}, \citenamefont {Guinea},\
  and\ \citenamefont {Castro~Neto}}]{Peres2006}%
  \BibitemOpen
  \bibfield  {author} {\bibinfo {author} {\bibfnamefont {N.~M.~R.}\
  \bibnamefont {Peres}}, \bibinfo {author} {\bibfnamefont {F.}~\bibnamefont
  {Guinea}},\ and\ \bibinfo {author} {\bibfnamefont {A.~H.}\ \bibnamefont
  {Castro~Neto}},\ }\bibfield  {title} {\bibinfo {title} {Electronic properties
  of disordered two-dimensional carbon},\ }\href
  {https://doi.org/10.1103/PhysRevB.73.125411} {\bibfield  {journal} {\bibinfo
  {journal} {Phys. Rev. B}\ }\textbf {\bibinfo {volume} {73}},\ \bibinfo
  {pages} {125411} (\bibinfo {year} {2006}{\natexlab{a}})}\BibitemShut
  {NoStop}%
\bibitem [{\citenamefont {Zhu}\ and\ \citenamefont {Heyl}(2021)}]{Zhu2021}%
  \BibitemOpen
  \bibfield  {author} {\bibinfo {author} {\bibfnamefont {G.-Y.}\ \bibnamefont
  {Zhu}}\ and\ \bibinfo {author} {\bibfnamefont {M.}~\bibnamefont {Heyl}},\
  }\bibfield  {title} {\bibinfo {title} {Subdiffusive dynamics and critical
  quantum correlations in a disorder-free localized {Kitaev} honeycomb model
  out of equilibrium},\ }\href
  {https://doi.org/10.1103/PhysRevResearch.3.L032069} {\bibfield  {journal}
  {\bibinfo  {journal} {Phys. Rev. Research}\ }\textbf {\bibinfo {volume}
  {3}},\ \bibinfo {pages} {L032069} (\bibinfo {year} {2021})}\BibitemShut
  {NoStop}%
\bibitem [{\citenamefont {Kim}\ \emph {et~al.}(2022)\citenamefont {Kim},
  \citenamefont {De~Tomasi},\ and\ \citenamefont {Castelnovo}}]{Kim2022}%
  \BibitemOpen
  \bibfield  {author} {\bibinfo {author} {\bibfnamefont {M.}~\bibnamefont
  {Kim}}, \bibinfo {author} {\bibfnamefont {G.}~\bibnamefont {De~Tomasi}},\
  and\ \bibinfo {author} {\bibfnamefont {C.}~\bibnamefont {Castelnovo}},\
  }\bibfield  {title} {\bibinfo {title} {{Anderson localization of emergent
  quasiparticles: Spinon and vison interplay at finite temperature in a
  ${\mathbb{Z}}_{2}$ gauge theory in three dimensions}},\ }\href
  {https://doi.org/10.1103/PhysRevResearch.4.043206} {\bibfield  {journal}
  {\bibinfo  {journal} {Phys. Rev. Res.}\ }\textbf {\bibinfo {volume} {4}},\
  \bibinfo {pages} {043206} (\bibinfo {year} {2022})}\BibitemShut {NoStop}%
\bibitem [{\citenamefont {Ohtsuki}\ \emph {et~al.}(1993)\citenamefont
  {Ohtsuki}, \citenamefont {Slevin},\ and\ \citenamefont {Ono}}]{Ohtsuki1993}%
  \BibitemOpen
  \bibfield  {author} {\bibinfo {author} {\bibfnamefont {T.}~\bibnamefont
  {Ohtsuki}}, \bibinfo {author} {\bibfnamefont {K.}~\bibnamefont {Slevin}},\
  and\ \bibinfo {author} {\bibfnamefont {Y.}~\bibnamefont {Ono}},\ }\bibfield
  {title} {\bibinfo {title} {Conductance fluctuations in two-dimensional
  systems in random magnetic fields},\ }\href
  {https://doi.org/10.1143/JPSJ.62.3979} {\bibfield  {journal} {\bibinfo
  {journal} {Journal of the Physical Society of Japan}\ }\textbf {\bibinfo
  {volume} {62}},\ \bibinfo {pages} {3979} (\bibinfo {year}
  {1993})}\BibitemShut {NoStop}%
\bibitem [{\citenamefont {Tadjine}\ and\ \citenamefont
  {Delerue}(2018)}]{Tadjine2018}%
  \BibitemOpen
  \bibfield  {author} {\bibinfo {author} {\bibfnamefont {A.}~\bibnamefont
  {Tadjine}}\ and\ \bibinfo {author} {\bibfnamefont {C.}~\bibnamefont
  {Delerue}},\ }\bibfield  {title} {\bibinfo {title} {Anderson localization
  induced by gauge-invariant bond-sign disorder in square $\mathrm{PbSe}$
  nanocrystal lattices},\ }\href {https://doi.org/10.1103/PhysRevB.98.125412}
  {\bibfield  {journal} {\bibinfo  {journal} {Phys. Rev. B}\ }\textbf {\bibinfo
  {volume} {98}},\ \bibinfo {pages} {125412} (\bibinfo {year}
  {2018})}\BibitemShut {NoStop}%
\bibitem [{\citenamefont {Hart}\ \emph {et~al.}(2020)\citenamefont {Hart},
  \citenamefont {Wan},\ and\ \citenamefont {Castelnovo}}]{Hart2020}%
  \BibitemOpen
  \bibfield  {author} {\bibinfo {author} {\bibfnamefont {O.}~\bibnamefont
  {Hart}}, \bibinfo {author} {\bibfnamefont {Y.}~\bibnamefont {Wan}},\ and\
  \bibinfo {author} {\bibfnamefont {C.}~\bibnamefont {Castelnovo}},\ }\bibfield
   {title} {\bibinfo {title} {Coherent propagation of quasiparticles in
  topological spin liquids at finite temperature},\ }\href
  {https://doi.org/10.1103/PhysRevB.101.064428} {\bibfield  {journal} {\bibinfo
   {journal} {Phys. Rev. B}\ }\textbf {\bibinfo {volume} {101}},\ \bibinfo
  {pages} {064428} (\bibinfo {year} {2020})}\BibitemShut {NoStop}%
\bibitem [{\citenamefont {Hart}\ \emph {et~al.}(2021)\citenamefont {Hart},
  \citenamefont {Wan},\ and\ \citenamefont {Castelnovo}}]{Hart2021}%
  \BibitemOpen
  \bibfield  {author} {\bibinfo {author} {\bibfnamefont {O.}~\bibnamefont
  {Hart}}, \bibinfo {author} {\bibfnamefont {Y.}~\bibnamefont {Wan}},\ and\
  \bibinfo {author} {\bibfnamefont {C.}~\bibnamefont {Castelnovo}},\ }\bibfield
   {title} {\bibinfo {title} {Correlation holes and slow dynamics induced by
  fractional statistics in gapped quantum spin liquids},\ }\href
  {https://doi.org/10.1038/s41467-021-21495-8} {\bibfield  {journal} {\bibinfo
  {journal} {Nature communications}\ }\textbf {\bibinfo {volume} {12}},\
  \bibinfo {pages} {1459} (\bibinfo {year} {2021})}\BibitemShut {NoStop}%
\bibitem [{\citenamefont {Li}\ \emph {et~al.}(2022)\citenamefont {Li},
  \citenamefont {Zhang}, \citenamefont {Budich},\ and\ \citenamefont
  {Trauzettel}}]{Li2022}%
  \BibitemOpen
  \bibfield  {author} {\bibinfo {author} {\bibfnamefont {C.-A.}\ \bibnamefont
  {Li}}, \bibinfo {author} {\bibfnamefont {S.-B.}\ \bibnamefont {Zhang}},
  \bibinfo {author} {\bibfnamefont {J.~C.}\ \bibnamefont {Budich}},\ and\
  \bibinfo {author} {\bibfnamefont {B.}~\bibnamefont {Trauzettel}},\ }\bibfield
   {title} {\bibinfo {title} {Transition from metal to higher-order topological
  insulator driven by random flux},\ }\href
  {https://doi.org/10.1103/PhysRevB.106.L081410} {\bibfield  {journal}
  {\bibinfo  {journal} {Phys. Rev. B}\ }\textbf {\bibinfo {volume} {106}},\
  \bibinfo {pages} {L081410} (\bibinfo {year} {2022})}\BibitemShut {NoStop}%
\bibitem [{\citenamefont {Altland}\ and\ \citenamefont
  {Zirnbauer}(1997)}]{Altland1997}%
  \BibitemOpen
  \bibfield  {author} {\bibinfo {author} {\bibfnamefont {A.}~\bibnamefont
  {Altland}}\ and\ \bibinfo {author} {\bibfnamefont {M.~R.}\ \bibnamefont
  {Zirnbauer}},\ }\bibfield  {title} {\bibinfo {title} {Nonstandard symmetry
  classes in mesoscopic normal-superconducting hybrid structures},\ }\href
  {https://doi.org/10.1103/PhysRevB.55.1142} {\bibfield  {journal} {\bibinfo
  {journal} {Phys. Rev. B}\ }\textbf {\bibinfo {volume} {55}},\ \bibinfo
  {pages} {1142} (\bibinfo {year} {1997})}\BibitemShut {NoStop}%
\bibitem [{\citenamefont {Evers}\ and\ \citenamefont
  {Mirlin}(2008)}]{Evers2008}%
  \BibitemOpen
  \bibfield  {author} {\bibinfo {author} {\bibfnamefont {F.}~\bibnamefont
  {Evers}}\ and\ \bibinfo {author} {\bibfnamefont {A.~D.}\ \bibnamefont
  {Mirlin}},\ }\bibfield  {title} {\bibinfo {title} {Anderson transitions},\
  }\href {https://doi.org/10.1103/RevModPhys.80.1355} {\bibfield  {journal}
  {\bibinfo  {journal} {Rev. Mod. Phys.}\ }\textbf {\bibinfo {volume} {80}},\
  \bibinfo {pages} {1355} (\bibinfo {year} {2008})}\BibitemShut {NoStop}%
\bibitem [{\citenamefont {Ludwig}(2015)}]{Ludwig2016}%
  \BibitemOpen
  \bibfield  {author} {\bibinfo {author} {\bibfnamefont {A.~W.~W.}\
  \bibnamefont {Ludwig}},\ }\bibfield  {title} {\bibinfo {title} {Topological
  phases: classification of topological insulators and superconductors of
  non-interacting fermions, and beyond},\ }\href
  {https://doi.org/10.1088/0031-8949/2015/T168/014001} {\bibfield  {journal}
  {\bibinfo  {journal} {Physica Scripta}\ }\textbf {\bibinfo {volume} {2016}},\
  \bibinfo {pages} {014001} (\bibinfo {year} {2015})}\BibitemShut {NoStop}%
\bibitem [{\citenamefont {Castro~Neto}\ \emph {et~al.}(2009)\citenamefont
  {Castro~Neto}, \citenamefont {Guinea}, \citenamefont {Peres}, \citenamefont
  {Novoselov},\ and\ \citenamefont {Geim}}]{Castro2009}%
  \BibitemOpen
  \bibfield  {author} {\bibinfo {author} {\bibfnamefont {A.~H.}\ \bibnamefont
  {Castro~Neto}}, \bibinfo {author} {\bibfnamefont {F.}~\bibnamefont {Guinea}},
  \bibinfo {author} {\bibfnamefont {N.~M.~R.}\ \bibnamefont {Peres}}, \bibinfo
  {author} {\bibfnamefont {K.~S.}\ \bibnamefont {Novoselov}},\ and\ \bibinfo
  {author} {\bibfnamefont {A.~K.}\ \bibnamefont {Geim}},\ }\bibfield  {title}
  {\bibinfo {title} {The electronic properties of graphene},\ }\href
  {https://doi.org/10.1103/RevModPhys.81.109} {\bibfield  {journal} {\bibinfo
  {journal} {Rev. Mod. Phys.}\ }\textbf {\bibinfo {volume} {81}},\ \bibinfo
  {pages} {109} (\bibinfo {year} {2009})}\BibitemShut {NoStop}%
\bibitem [{\citenamefont {Gade}\ and\ \citenamefont {Wegner}(1991)}]{Gade1991}%
  \BibitemOpen
  \bibfield  {author} {\bibinfo {author} {\bibfnamefont {R.}~\bibnamefont
  {Gade}}\ and\ \bibinfo {author} {\bibfnamefont {F.}~\bibnamefont {Wegner}},\
  }\bibfield  {title} {\bibinfo {title} {{The $n = 0$ replica limit of U($n$)
  and U($n$)SO($n$) models}},\ }\href
  {https://doi.org/https://doi.org/10.1016/0550-3213(91)90401-I} {\bibfield
  {journal} {\bibinfo  {journal} {Nuclear Physics B}\ }\textbf {\bibinfo
  {volume} {360}},\ \bibinfo {pages} {213} (\bibinfo {year}
  {1991})}\BibitemShut {NoStop}%
\bibitem [{\citenamefont {Gade}(1993)}]{Gade1993}%
  \BibitemOpen
  \bibfield  {author} {\bibinfo {author} {\bibfnamefont {R.}~\bibnamefont
  {Gade}},\ }\bibfield  {title} {\bibinfo {title} {Anderson localization for
  sublattice models},\ }\href
  {https://doi.org/https://doi.org/10.1016/0550-3213(93)90601-K} {\bibfield
  {journal} {\bibinfo  {journal} {Nuclear Physics B}\ }\textbf {\bibinfo
  {volume} {398}},\ \bibinfo {pages} {499} (\bibinfo {year}
  {1993})}\BibitemShut {NoStop}%
\bibitem [{\citenamefont {H\"afner}\ \emph {et~al.}(2014)\citenamefont
  {H\"afner}, \citenamefont {Schindler}, \citenamefont {Weik}, \citenamefont
  {Mayer}, \citenamefont {Balakrishnan}, \citenamefont {Narayanan},
  \citenamefont {Bera},\ and\ \citenamefont {Evers}}]{Hafner2014}%
  \BibitemOpen
  \bibfield  {author} {\bibinfo {author} {\bibfnamefont {V.}~\bibnamefont
  {H\"afner}}, \bibinfo {author} {\bibfnamefont {J.}~\bibnamefont {Schindler}},
  \bibinfo {author} {\bibfnamefont {N.}~\bibnamefont {Weik}}, \bibinfo {author}
  {\bibfnamefont {T.}~\bibnamefont {Mayer}}, \bibinfo {author} {\bibfnamefont
  {S.}~\bibnamefont {Balakrishnan}}, \bibinfo {author} {\bibfnamefont
  {R.}~\bibnamefont {Narayanan}}, \bibinfo {author} {\bibfnamefont
  {S.}~\bibnamefont {Bera}},\ and\ \bibinfo {author} {\bibfnamefont
  {F.}~\bibnamefont {Evers}},\ }\bibfield  {title} {\bibinfo {title} {Density
  of states in graphene with vacancies: Midgap power law and frozen
  multifractality},\ }\href {https://doi.org/10.1103/PhysRevLett.113.186802}
  {\bibfield  {journal} {\bibinfo  {journal} {Phys. Rev. Lett.}\ }\textbf
  {\bibinfo {volume} {113}},\ \bibinfo {pages} {186802} (\bibinfo {year}
  {2014})}\BibitemShut {NoStop}%
\bibitem [{\citenamefont {Ferreira}\ and\ \citenamefont
  {Mucciolo}(2015)}]{Ferreira2015}%
  \BibitemOpen
  \bibfield  {author} {\bibinfo {author} {\bibfnamefont {A.}~\bibnamefont
  {Ferreira}}\ and\ \bibinfo {author} {\bibfnamefont {E.~R.}\ \bibnamefont
  {Mucciolo}},\ }\bibfield  {title} {\bibinfo {title} {Critical delocalization
  of chiral zero energy modes in graphene},\ }\href
  {https://doi.org/10.1103/PhysRevLett.115.106601} {\bibfield  {journal}
  {\bibinfo  {journal} {Phys. Rev. Lett.}\ }\textbf {\bibinfo {volume} {115}},\
  \bibinfo {pages} {106601} (\bibinfo {year} {2015})}\BibitemShut {NoStop}%
\bibitem [{\citenamefont {Sanyal}\ \emph {et~al.}(2016)\citenamefont {Sanyal},
  \citenamefont {Damle},\ and\ \citenamefont {Motrunich}}]{Sanyal2016}%
  \BibitemOpen
  \bibfield  {author} {\bibinfo {author} {\bibfnamefont {S.}~\bibnamefont
  {Sanyal}}, \bibinfo {author} {\bibfnamefont {K.}~\bibnamefont {Damle}},\ and\
  \bibinfo {author} {\bibfnamefont {O.~I.}\ \bibnamefont {Motrunich}},\
  }\bibfield  {title} {\bibinfo {title} {Vacancy-induced low-energy states in
  undoped graphene},\ }\href {https://doi.org/10.1103/PhysRevLett.117.116806}
  {\bibfield  {journal} {\bibinfo  {journal} {Phys. Rev. Lett.}\ }\textbf
  {\bibinfo {volume} {117}},\ \bibinfo {pages} {116806} (\bibinfo {year}
  {2016})}\BibitemShut {NoStop}%
\bibitem [{\citenamefont {Ostrovsky}\ \emph {et~al.}(2014)\citenamefont
  {Ostrovsky}, \citenamefont {Protopopov}, \citenamefont {K\"onig},
  \citenamefont {Gornyi}, \citenamefont {Mirlin},\ and\ \citenamefont
  {Skvortsov}}]{Ostrovsky2014}%
  \BibitemOpen
  \bibfield  {author} {\bibinfo {author} {\bibfnamefont {P.~M.}\ \bibnamefont
  {Ostrovsky}}, \bibinfo {author} {\bibfnamefont {I.~V.}\ \bibnamefont
  {Protopopov}}, \bibinfo {author} {\bibfnamefont {E.~J.}\ \bibnamefont
  {K\"onig}}, \bibinfo {author} {\bibfnamefont {I.~V.}\ \bibnamefont {Gornyi}},
  \bibinfo {author} {\bibfnamefont {A.~D.}\ \bibnamefont {Mirlin}},\ and\
  \bibinfo {author} {\bibfnamefont {M.~A.}\ \bibnamefont {Skvortsov}},\
  }\bibfield  {title} {\bibinfo {title} {Density of states in a two-dimensional
  chiral metal with vacancies},\ }\href
  {https://doi.org/10.1103/PhysRevLett.113.186803} {\bibfield  {journal}
  {\bibinfo  {journal} {Phys. Rev. Lett.}\ }\textbf {\bibinfo {volume} {113}},\
  \bibinfo {pages} {186803} (\bibinfo {year} {2014})}\BibitemShut {NoStop}%
\bibitem [{\citenamefont {Titov}(2007)}]{Titov2007}%
  \BibitemOpen
  \bibfield  {author} {\bibinfo {author} {\bibfnamefont {M.}~\bibnamefont
  {Titov}},\ }\bibfield  {title} {\bibinfo {title} {Impurity-assisted tunneling
  in graphene},\ }\href {https://doi.org/10.1209/0295-5075/79/17004} {\bibfield
   {journal} {\bibinfo  {journal} {Europhysics Letters}\ }\textbf {\bibinfo
  {volume} {79}},\ \bibinfo {pages} {17004} (\bibinfo {year}
  {2007})}\BibitemShut {NoStop}%
\bibitem [{\citenamefont {Brey}\ and\ \citenamefont {Fertig}(2006)}]{Brey2006}%
  \BibitemOpen
  \bibfield  {author} {\bibinfo {author} {\bibfnamefont {L.}~\bibnamefont
  {Brey}}\ and\ \bibinfo {author} {\bibfnamefont {H.~A.}\ \bibnamefont
  {Fertig}},\ }\bibfield  {title} {\bibinfo {title} {Electronic states of
  graphene nanoribbons studied with the {Dirac} equation},\ }\href
  {https://doi.org/10.1103/PhysRevB.73.235411} {\bibfield  {journal} {\bibinfo
  {journal} {Phys. Rev. B}\ }\textbf {\bibinfo {volume} {73}},\ \bibinfo
  {pages} {235411} (\bibinfo {year} {2006})}\BibitemShut {NoStop}%
\bibitem [{\citenamefont {Peres}\ \emph
  {et~al.}(2006{\natexlab{b}})\citenamefont {Peres}, \citenamefont
  {Castro~Neto},\ and\ \citenamefont {Guinea}}]{Peres2006_2}%
  \BibitemOpen
  \bibfield  {author} {\bibinfo {author} {\bibfnamefont {N.~M.~R.}\
  \bibnamefont {Peres}}, \bibinfo {author} {\bibfnamefont {A.~H.}\ \bibnamefont
  {Castro~Neto}},\ and\ \bibinfo {author} {\bibfnamefont {F.}~\bibnamefont
  {Guinea}},\ }\bibfield  {title} {\bibinfo {title} {Conductance quantization
  in mesoscopic graphene},\ }\href {https://doi.org/10.1103/PhysRevB.73.195411}
  {\bibfield  {journal} {\bibinfo  {journal} {Phys. Rev. B}\ }\textbf {\bibinfo
  {volume} {73}},\ \bibinfo {pages} {195411} (\bibinfo {year}
  {2006}{\natexlab{b}})}\BibitemShut {NoStop}%
\bibitem [{\citenamefont {Landauer}(1992)}]{Landauer1992}%
  \BibitemOpen
  \bibfield  {author} {\bibinfo {author} {\bibfnamefont {R.}~\bibnamefont
  {Landauer}},\ }\bibfield  {title} {\bibinfo {title} {Conductance from
  transmission: common sense points},\ }\href
  {https://doi.org/10.1088/0031-8949/1992/T42/020} {\bibfield  {journal}
  {\bibinfo  {journal} {Physica Scripta}\ }\textbf {\bibinfo {volume} {1992}},\
  \bibinfo {pages} {110} (\bibinfo {year} {1992})}\BibitemShut {NoStop}%
\bibitem [{\citenamefont {Fisher}\ and\ \citenamefont
  {Lee}(1981)}]{Fisher1981}%
  \BibitemOpen
  \bibfield  {author} {\bibinfo {author} {\bibfnamefont {D.~S.}\ \bibnamefont
  {Fisher}}\ and\ \bibinfo {author} {\bibfnamefont {P.~A.}\ \bibnamefont
  {Lee}},\ }\bibfield  {title} {\bibinfo {title} {Relation between conductivity
  and transmission matrix},\ }\href {https://doi.org/10.1103/PhysRevB.23.6851}
  {\bibfield  {journal} {\bibinfo  {journal} {Phys. Rev. B}\ }\textbf {\bibinfo
  {volume} {23}},\ \bibinfo {pages} {6851} (\bibinfo {year}
  {1981})}\BibitemShut {NoStop}%
\bibitem [{\citenamefont {Meir}\ and\ \citenamefont
  {Wingreen}(1992)}]{Meir1992}%
  \BibitemOpen
  \bibfield  {author} {\bibinfo {author} {\bibfnamefont {Y.}~\bibnamefont
  {Meir}}\ and\ \bibinfo {author} {\bibfnamefont {N.~S.}\ \bibnamefont
  {Wingreen}},\ }\bibfield  {title} {\bibinfo {title} {Landauer formula for the
  current through an interacting electron region},\ }\href
  {https://doi.org/10.1103/PhysRevLett.68.2512} {\bibfield  {journal} {\bibinfo
   {journal} {Phys. Rev. Lett.}\ }\textbf {\bibinfo {volume} {68}},\ \bibinfo
  {pages} {2512} (\bibinfo {year} {1992})}\BibitemShut {NoStop}%
\bibitem [{\citenamefont {Sancho}\ \emph {et~al.}(1984)\citenamefont {Sancho},
  \citenamefont {Sancho},\ and\ \citenamefont {Rubio}}]{Sancho1984}%
  \BibitemOpen
  \bibfield  {author} {\bibinfo {author} {\bibfnamefont {M.~P.~L.}\
  \bibnamefont {Sancho}}, \bibinfo {author} {\bibfnamefont {J.~M.~L.}\
  \bibnamefont {Sancho}},\ and\ \bibinfo {author} {\bibfnamefont
  {J.}~\bibnamefont {Rubio}},\ }\bibfield  {title} {\bibinfo {title} {Quick
  iterative scheme for the calculation of transfer matrices: application to
  {Mo} (100)},\ }\href {https://doi.org/10.1088/0305-4608/14/5/016} {\bibfield
  {journal} {\bibinfo  {journal} {Journal of Physics F: Metal Physics}\
  }\textbf {\bibinfo {volume} {14}},\ \bibinfo {pages} {1205} (\bibinfo {year}
  {1984})}\BibitemShut {NoStop}%
\bibitem [{\citenamefont {Nardelli}(1999)}]{Nardelli1999}%
  \BibitemOpen
  \bibfield  {author} {\bibinfo {author} {\bibfnamefont {M.~B.}\ \bibnamefont
  {Nardelli}},\ }\bibfield  {title} {\bibinfo {title} {Electronic transport in
  extended systems: Application to carbon nanotubes},\ }\href
  {https://doi.org/10.1103/PhysRevB.60.7828} {\bibfield  {journal} {\bibinfo
  {journal} {Phys. Rev. B}\ }\textbf {\bibinfo {volume} {60}},\ \bibinfo
  {pages} {7828} (\bibinfo {year} {1999})}\BibitemShut {NoStop}%
\bibitem [{\citenamefont {MacKinnon}(1985)}]{Mackinnon1985}%
  \BibitemOpen
  \bibfield  {author} {\bibinfo {author} {\bibfnamefont {A.}~\bibnamefont
  {MacKinnon}},\ }\bibfield  {title} {\bibinfo {title} {The calculation of
  transport properties and density of states of disordered solids},\ }\href
  {https://doi.org/10.1007/BF01328846} {\bibfield  {journal} {\bibinfo
  {journal} {Zeitschrift f{\"u}r Physik B Condensed Matter}\ }\textbf {\bibinfo
  {volume} {59}},\ \bibinfo {pages} {385} (\bibinfo {year} {1985})}\BibitemShut
  {NoStop}%
\bibitem [{\citenamefont {Croy}\ \emph {et~al.}(2006)\citenamefont {Croy},
  \citenamefont {R{\"o}mer},\ and\ \citenamefont {Schreiber}}]{Croy2006}%
  \BibitemOpen
  \bibfield  {author} {\bibinfo {author} {\bibfnamefont {A.}~\bibnamefont
  {Croy}}, \bibinfo {author} {\bibfnamefont {R.~A.}\ \bibnamefont
  {R{\"o}mer}},\ and\ \bibinfo {author} {\bibfnamefont {M.}~\bibnamefont
  {Schreiber}},\ }\bibfield  {title} {\bibinfo {title} {{Localization of
  Electronic States in Amorphous Materials: Recursive Green's Function Method
  and the Metal-Insulator Transition at $E \neq 0$}},\ }in\ \href@noop {}
  {\emph {\bibinfo {booktitle} {Parallel Algorithms and Cluster Computing}}},\
  \bibinfo {editor} {edited by\ \bibinfo {editor} {\bibfnamefont {K.~H.}\
  \bibnamefont {Hoffmann}}\ and\ \bibinfo {editor} {\bibfnamefont
  {A.}~\bibnamefont {Meyer}}}\ (\bibinfo  {publisher} {Springer Berlin
  Heidelberg},\ \bibinfo {address} {Berlin, Heidelberg},\ \bibinfo {year}
  {2006})\ pp.\ \bibinfo {pages} {203--226}\BibitemShut {NoStop}%
\bibitem [{\citenamefont {Das~Sarma}\ \emph {et~al.}(2011)\citenamefont
  {Das~Sarma}, \citenamefont {Adam}, \citenamefont {Hwang},\ and\ \citenamefont
  {Rossi}}]{DasSarma2011}%
  \BibitemOpen
  \bibfield  {author} {\bibinfo {author} {\bibfnamefont {S.}~\bibnamefont
  {Das~Sarma}}, \bibinfo {author} {\bibfnamefont {S.}~\bibnamefont {Adam}},
  \bibinfo {author} {\bibfnamefont {E.~H.}\ \bibnamefont {Hwang}},\ and\
  \bibinfo {author} {\bibfnamefont {E.}~\bibnamefont {Rossi}},\ }\bibfield
  {title} {\bibinfo {title} {Electronic transport in two-dimensional
  graphene},\ }\href {https://doi.org/10.1103/RevModPhys.83.407} {\bibfield
  {journal} {\bibinfo  {journal} {Rev. Mod. Phys.}\ }\textbf {\bibinfo {volume}
  {83}},\ \bibinfo {pages} {407} (\bibinfo {year} {2011})}\BibitemShut
  {NoStop}%
\bibitem [{\citenamefont {Peres}(2010)}]{Peres2010}%
  \BibitemOpen
  \bibfield  {author} {\bibinfo {author} {\bibfnamefont {N.~M.~R.}\
  \bibnamefont {Peres}},\ }\bibfield  {title} {\bibinfo {title} {Colloquium:
  The transport properties of graphene: An introduction},\ }\href
  {https://doi.org/10.1103/RevModPhys.82.2673} {\bibfield  {journal} {\bibinfo
  {journal} {Rev. Mod. Phys.}\ }\textbf {\bibinfo {volume} {82}},\ \bibinfo
  {pages} {2673} (\bibinfo {year} {2010})}\BibitemShut {NoStop}%
\bibitem [{\citenamefont {Lee}\ and\ \citenamefont
  {Ramakrishnan}(1985)}]{Lee1985}%
  \BibitemOpen
  \bibfield  {author} {\bibinfo {author} {\bibfnamefont {P.~A.}\ \bibnamefont
  {Lee}}\ and\ \bibinfo {author} {\bibfnamefont {T.~V.}\ \bibnamefont
  {Ramakrishnan}},\ }\bibfield  {title} {\bibinfo {title} {Disordered
  electronic systems},\ }\href {https://doi.org/10.1103/RevModPhys.57.287}
  {\bibfield  {journal} {\bibinfo  {journal} {Rev. Mod. Phys.}\ }\textbf
  {\bibinfo {volume} {57}},\ \bibinfo {pages} {287} (\bibinfo {year}
  {1985})}\BibitemShut {NoStop}%
\bibitem [{\citenamefont {MacKinnon}\ and\ \citenamefont
  {Kramer}(1981)}]{Mackinnon1981}%
  \BibitemOpen
  \bibfield  {author} {\bibinfo {author} {\bibfnamefont {A.}~\bibnamefont
  {MacKinnon}}\ and\ \bibinfo {author} {\bibfnamefont {B.}~\bibnamefont
  {Kramer}},\ }\bibfield  {title} {\bibinfo {title} {One-parameter scaling of
  localization length and conductance in disordered systems},\ }\href
  {https://doi.org/10.1103/PhysRevLett.47.1546} {\bibfield  {journal} {\bibinfo
   {journal} {Phys. Rev. Lett.}\ }\textbf {\bibinfo {volume} {47}},\ \bibinfo
  {pages} {1546} (\bibinfo {year} {1981})}\BibitemShut {NoStop}%
\bibitem [{\citenamefont {MacKinnon}\ and\ \citenamefont
  {Kramer}(1983)}]{MacKinnon1983}%
  \BibitemOpen
  \bibfield  {author} {\bibinfo {author} {\bibfnamefont {A.}~\bibnamefont
  {MacKinnon}}\ and\ \bibinfo {author} {\bibfnamefont {B.}~\bibnamefont
  {Kramer}},\ }\bibfield  {title} {\bibinfo {title} {The scaling theory of
  electrons in disordered solids: Additional numerical results},\ }\href
  {https://doi.org/10.1007/BF01578242} {\bibfield  {journal} {\bibinfo
  {journal} {Zeitschrift f{\"u}r Physik B Condensed Matter}\ }\textbf {\bibinfo
  {volume} {53}},\ \bibinfo {pages} {1} (\bibinfo {year} {1983})}\BibitemShut
  {NoStop}%
\bibitem [{\citenamefont {Nakada}\ \emph {et~al.}(1996)\citenamefont {Nakada},
  \citenamefont {Fujita}, \citenamefont {Dresselhaus},\ and\ \citenamefont
  {Dresselhaus}}]{Nakada1996}%
  \BibitemOpen
  \bibfield  {author} {\bibinfo {author} {\bibfnamefont {K.}~\bibnamefont
  {Nakada}}, \bibinfo {author} {\bibfnamefont {M.}~\bibnamefont {Fujita}},
  \bibinfo {author} {\bibfnamefont {G.}~\bibnamefont {Dresselhaus}},\ and\
  \bibinfo {author} {\bibfnamefont {M.~S.}\ \bibnamefont {Dresselhaus}},\
  }\bibfield  {title} {\bibinfo {title} {Edge state in graphene ribbons:
  Nanometer size effect and edge shape dependence},\ }\href
  {https://doi.org/10.1103/PhysRevB.54.17954} {\bibfield  {journal} {\bibinfo
  {journal} {Phys. Rev. B}\ }\textbf {\bibinfo {volume} {54}},\ \bibinfo
  {pages} {17954} (\bibinfo {year} {1996})}\BibitemShut {NoStop}%
\bibitem [{\citenamefont {Marko\v{s}}\ and\ \citenamefont
  {Schweitzer}(2007)}]{Markos2007}%
  \BibitemOpen
  \bibfield  {author} {\bibinfo {author} {\bibfnamefont {P.}~\bibnamefont
  {Marko\v{s}}}\ and\ \bibinfo {author} {\bibfnamefont {L.}~\bibnamefont
  {Schweitzer}},\ }\bibfield  {title} {\bibinfo {title} {Critical conductance
  of two-dimensional chiral systems with random magnetic flux},\ }\href
  {https://doi.org/10.1103/PhysRevB.76.115318} {\bibfield  {journal} {\bibinfo
  {journal} {Phys. Rev. B}\ }\textbf {\bibinfo {volume} {76}},\ \bibinfo
  {pages} {115318} (\bibinfo {year} {2007})}\BibitemShut {NoStop}%
\bibitem [{\citenamefont {Furusaki}(1999)}]{Furusaki1999}%
  \BibitemOpen
  \bibfield  {author} {\bibinfo {author} {\bibfnamefont {A.}~\bibnamefont
  {Furusaki}},\ }\bibfield  {title} {\bibinfo {title} {Anderson localization
  due to a random magnetic field in two dimensions},\ }\href
  {https://doi.org/10.1103/PhysRevLett.82.604} {\bibfield  {journal} {\bibinfo
  {journal} {Phys. Rev. Lett.}\ }\textbf {\bibinfo {volume} {82}},\ \bibinfo
  {pages} {604} (\bibinfo {year} {1999})}\BibitemShut {NoStop}%
\bibitem [{\citenamefont {Schweitzer}\ and\ \citenamefont
  {Markoš}(2008)}]{Schweitzer2008}%
  \BibitemOpen
  \bibfield  {author} {\bibinfo {author} {\bibfnamefont {L.}~\bibnamefont
  {Schweitzer}}\ and\ \bibinfo {author} {\bibfnamefont {P.}~\bibnamefont
  {Markoš}},\ }\bibfield  {title} {\bibinfo {title} {Critical conductance of
  the chiral two-dimensional random flux model},\ }\href
  {https://doi.org/https://doi.org/10.1016/j.physe.2007.08.083} {\bibfield
  {journal} {\bibinfo  {journal} {Physica E: Low-dimensional Systems and
  Nanostructures}\ }\textbf {\bibinfo {volume} {40}},\ \bibinfo {pages} {1335}
  (\bibinfo {year} {2008})}\BibitemShut {NoStop}%
\bibitem [{\citenamefont {Hatsugai}\ \emph {et~al.}(1997)\citenamefont
  {Hatsugai}, \citenamefont {Wen},\ and\ \citenamefont
  {Kohmoto}}]{Hatsugai1997}%
  \BibitemOpen
  \bibfield  {author} {\bibinfo {author} {\bibfnamefont {Y.}~\bibnamefont
  {Hatsugai}}, \bibinfo {author} {\bibfnamefont {X.-G.}\ \bibnamefont {Wen}},\
  and\ \bibinfo {author} {\bibfnamefont {M.}~\bibnamefont {Kohmoto}},\
  }\bibfield  {title} {\bibinfo {title} {Disordered critical wave functions in
  random-bond models in two dimensions: Random-lattice fermions at {$E=0$}
  without doubling},\ }\href {https://doi.org/10.1103/PhysRevB.56.1061}
  {\bibfield  {journal} {\bibinfo  {journal} {Phys. Rev. B}\ }\textbf {\bibinfo
  {volume} {56}},\ \bibinfo {pages} {1061} (\bibinfo {year}
  {1997})}\BibitemShut {NoStop}%
\bibitem [{\citenamefont {Lieb}(1994)}]{Lieb1994}%
  \BibitemOpen
  \bibfield  {author} {\bibinfo {author} {\bibfnamefont {E.~H.}\ \bibnamefont
  {Lieb}},\ }\bibfield  {title} {\bibinfo {title} {Flux phase of the
  half-filled band},\ }\href {https://doi.org/10.1103/PhysRevLett.73.2158}
  {\bibfield  {journal} {\bibinfo  {journal} {Phys. Rev. Lett.}\ }\textbf
  {\bibinfo {volume} {73}},\ \bibinfo {pages} {2158} (\bibinfo {year}
  {1994})}\BibitemShut {NoStop}%
\bibitem [{\citenamefont {Saito}\ \emph {et~al.}(2007)\citenamefont {Saito},
  \citenamefont {Nakamura},\ and\ \citenamefont {Natori}}]{Saito2007}%
  \BibitemOpen
  \bibfield  {author} {\bibinfo {author} {\bibfnamefont {K.}~\bibnamefont
  {Saito}}, \bibinfo {author} {\bibfnamefont {J.}~\bibnamefont {Nakamura}},\
  and\ \bibinfo {author} {\bibfnamefont {A.}~\bibnamefont {Natori}},\
  }\bibfield  {title} {\bibinfo {title} {Ballistic thermal conductance of a
  graphene sheet},\ }\href {https://doi.org/10.1103/PhysRevB.76.115409}
  {\bibfield  {journal} {\bibinfo  {journal} {Phys. Rev. B}\ }\textbf {\bibinfo
  {volume} {76}},\ \bibinfo {pages} {115409} (\bibinfo {year}
  {2007})}\BibitemShut {NoStop}%
\bibitem [{\citenamefont {Rycerz}(2021)}]{Rycerz2021}%
  \BibitemOpen
  \bibfield  {author} {\bibinfo {author} {\bibfnamefont {A.}~\bibnamefont
  {Rycerz}},\ }\bibfield  {title} {\bibinfo {title} {Wiedemann--franz law for
  massless {Dirac} fermions with implications for graphene},\ }\href
  {https://doi.org/10.3390/ma14112704} {\bibfield  {journal} {\bibinfo
  {journal} {Materials}\ }\textbf {\bibinfo {volume} {14}},\ \bibinfo {pages}
  {2704} (\bibinfo {year} {2021})}\BibitemShut {NoStop}%
\bibitem [{\citenamefont {Ponomarenko}\ \emph {et~al.}(2011)\citenamefont
  {Ponomarenko}, \citenamefont {Geim}, \citenamefont {Zhukov}, \citenamefont
  {Jalil}, \citenamefont {Morozov}, \citenamefont {Novoselov}, \citenamefont
  {Grigorieva}, \citenamefont {Hill}, \citenamefont {Cheianov}, \citenamefont
  {Fal’Ko} \emph {et~al.}}]{Ponomarenko2011}%
  \BibitemOpen
  \bibfield  {author} {\bibinfo {author} {\bibfnamefont {L.}~\bibnamefont
  {Ponomarenko}}, \bibinfo {author} {\bibfnamefont {A.}~\bibnamefont {Geim}},
  \bibinfo {author} {\bibfnamefont {A.}~\bibnamefont {Zhukov}}, \bibinfo
  {author} {\bibfnamefont {R.}~\bibnamefont {Jalil}}, \bibinfo {author}
  {\bibfnamefont {S.}~\bibnamefont {Morozov}}, \bibinfo {author} {\bibfnamefont
  {K.}~\bibnamefont {Novoselov}}, \bibinfo {author} {\bibfnamefont
  {I.}~\bibnamefont {Grigorieva}}, \bibinfo {author} {\bibfnamefont
  {E.}~\bibnamefont {Hill}}, \bibinfo {author} {\bibfnamefont {V.}~\bibnamefont
  {Cheianov}}, \bibinfo {author} {\bibfnamefont {V.}~\bibnamefont {Fal’Ko}},
  \emph {et~al.},\ }\bibfield  {title} {\bibinfo {title} {Tunable
  metal--insulator transition in double-layer graphene heterostructures},\
  }\href {https://doi.org/10.1038/nphys2114} {\bibfield  {journal} {\bibinfo
  {journal} {Nature Physics}\ }\textbf {\bibinfo {volume} {7}},\ \bibinfo
  {pages} {958} (\bibinfo {year} {2011})}\BibitemShut {NoStop}%
\bibitem [{\citenamefont {Moser}\ \emph {et~al.}(2010)\citenamefont {Moser},
  \citenamefont {Tao}, \citenamefont {Roche}, \citenamefont {Alzina},
  \citenamefont {Sotomayor~Torres},\ and\ \citenamefont
  {Bachtold}}]{Moser2010}%
  \BibitemOpen
  \bibfield  {author} {\bibinfo {author} {\bibfnamefont {J.}~\bibnamefont
  {Moser}}, \bibinfo {author} {\bibfnamefont {H.}~\bibnamefont {Tao}}, \bibinfo
  {author} {\bibfnamefont {S.}~\bibnamefont {Roche}}, \bibinfo {author}
  {\bibfnamefont {F.}~\bibnamefont {Alzina}}, \bibinfo {author} {\bibfnamefont
  {C.~M.}\ \bibnamefont {Sotomayor~Torres}},\ and\ \bibinfo {author}
  {\bibfnamefont {A.}~\bibnamefont {Bachtold}},\ }\bibfield  {title} {\bibinfo
  {title} {Magnetotransport in disordered graphene exposed to ozone: From weak
  to strong localization},\ }\href {https://doi.org/10.1103/PhysRevB.81.205445}
  {\bibfield  {journal} {\bibinfo  {journal} {Phys. Rev. B}\ }\textbf {\bibinfo
  {volume} {81}},\ \bibinfo {pages} {205445} (\bibinfo {year}
  {2010})}\BibitemShut {NoStop}%
\bibitem [{\citenamefont {Yanik}\ \emph {et~al.}(2021)\citenamefont {Yanik},
  \citenamefont {Sazgari}, \citenamefont {Canatar}, \citenamefont {Vaheb},\
  and\ \citenamefont {Kaya}}]{Yanik2021}%
  \BibitemOpen
  \bibfield  {author} {\bibinfo {author} {\bibfnamefont {C.}~\bibnamefont
  {Yanik}}, \bibinfo {author} {\bibfnamefont {V.}~\bibnamefont {Sazgari}},
  \bibinfo {author} {\bibfnamefont {A.}~\bibnamefont {Canatar}}, \bibinfo
  {author} {\bibfnamefont {Y.}~\bibnamefont {Vaheb}},\ and\ \bibinfo {author}
  {\bibfnamefont {I.~I.}\ \bibnamefont {Kaya}},\ }\bibfield  {title} {\bibinfo
  {title} {Strong localization in suspended monolayer graphene by intervalley
  scattering},\ }\href {https://doi.org/10.1103/PhysRevB.103.085437} {\bibfield
   {journal} {\bibinfo  {journal} {Phys. Rev. B}\ }\textbf {\bibinfo {volume}
  {103}},\ \bibinfo {pages} {085437} (\bibinfo {year} {2021})}\BibitemShut
  {NoStop}%
\end{thebibliography}%

\end{document}


\title{Supplemental Material for Transport in honeycomb lattice with random $\pi$-fluxes: implications for low-temperature thermal transport in the Kitaev spin liquids}
\author{Zekun Zhuang}
\affiliation{Center for Materials Theory, Rutgers University, Piscataway, New Jersey 08854, USA}
\maketitle

\section{Calculation of transmission coefficient using the recursive Green's function method}
\section{Calculation of localization length using the transfer matrix method}
To study the effects of localization in the RPFHM, we use the transfer matrix method \cite{Mackinnon1981,MacKinnon1983} to compute the Lyapunov exponent of the system around its Dirac point. We consider the same geometry shown in Fig. 3 in the main text, in which the system can be divided into successive slices labeled by $n$. The Schr\"odinger equation at given energy $E$ can be written in the form of
\begin{equation}
    \left(\begin{array}{c}
         |\Psi_{n+1}\rangle\\
          |\Psi_n\rangle
    \end{array}\right)=T_n    \left(\begin{array}{c}
         |\Psi_{n}\rangle\\
          |\Psi_{n-1}\rangle
    \end{array}\right),
\end{equation}
where $|\Psi_n\rangle$ is the wavefunction of slice $n$, $T_n$ is the transfer matrix
\begin{equation}
    T_n=\left(\begin{array}{cc}
       H_{n,n+1}^{-1}(E-H_{n,n})  &  -H_{n,n+1}^{-1} H_{n,n-1} \\
        \mathds{1} & 0
    \end{array}\right),
\end{equation}
and $H_{m,n}$ is the Hamiltonian matrix between slice $m$ and $n$. By iteration, one obtains
\begin{equation}
        \left(\begin{array}{c}
         |\Psi_{n+1}\rangle\\
          |\Psi_n\rangle
    \end{array}\right)=M_n    \left(\begin{array}{c}
         |\Psi_{1}\rangle\\
          |\Psi_{0}\rangle
    \end{array}\right),
\end{equation}
where $M_n=T_nT_{n-1}\cdots T_2T_1$. There exists a limiting matrix $M_\infty=\text{lim}_{n\rightarrow\infty}(M_nM_n^\dagger)^{1/(2n)}$, which has eigenvalues $e^{\gamma_i}$, where $\gamma_i$ is the Lyapunov exponent. The Lyapunov exponents must come in opposite pairs and the quasi-one-dimensional localization length $\lambda_M$ can be defined as the inverse of the smallest positive Lyapunov exponent. As the smallest Lyapunov exponent is susceptible to numerical error, we implement Gram-Schmidt orthonormalization every 8 steps for numerical stability (e.g. see \cite{Tadjine2018} and references herein).

\bibliography{Reference}